\documentclass[sigconf]{acmart}

\usepackage{amssymb}
\usepackage{amsmath}
\usepackage{booktabs}
\usepackage{mathtools}
\usepackage{pdfpages}
\usepackage{subcaption}
\usepackage{color}
\usepackage{multirow}
\usepackage{placeins}
\usepackage{setspace}

\usepackage{pifont}
\newcommand{\yescheck}{\ding{51}}
\newcommand{\nocheck}{\ding{55}}
\newcommand{\one}{\ding{182}}
\newcommand{\two}{\ding{183}}
\newcommand{\three}{\ding{184}}
\newcommand{\four}{\ding{185}}

\newcommand{\pages}[1]{}

\newcommand{\yin}{\includegraphics[height=0.8em]{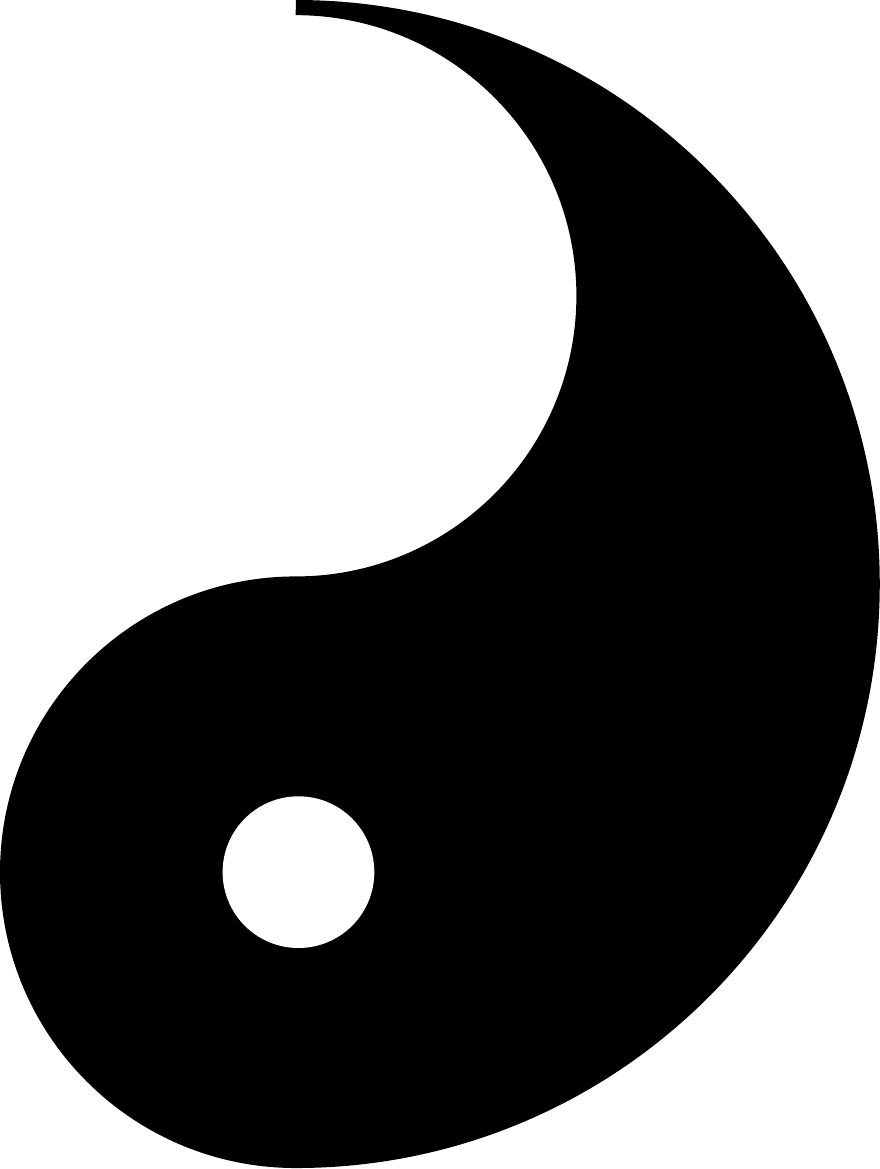}}
\newcommand{\yang}{\includegraphics[height=0.8em]{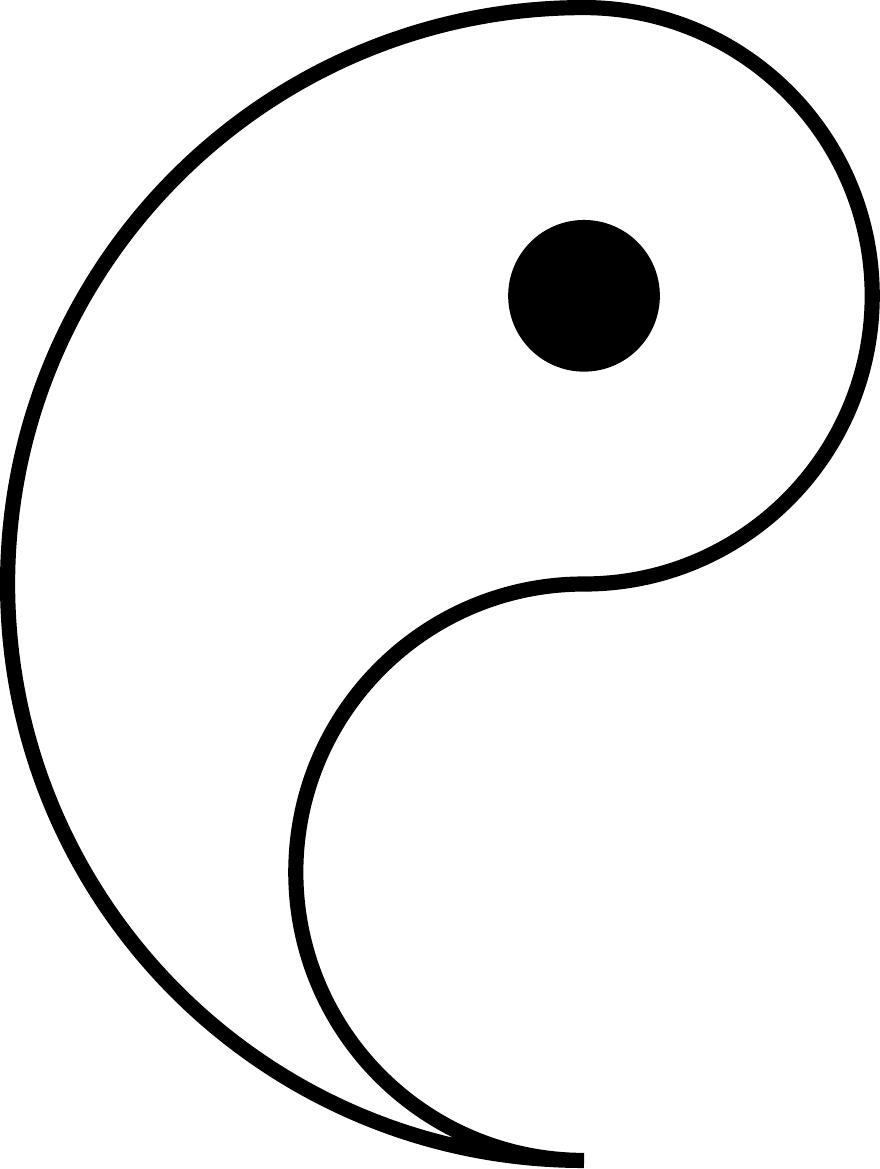}}
\newcommand{\yinyang}{\includegraphics[height=0.8em]{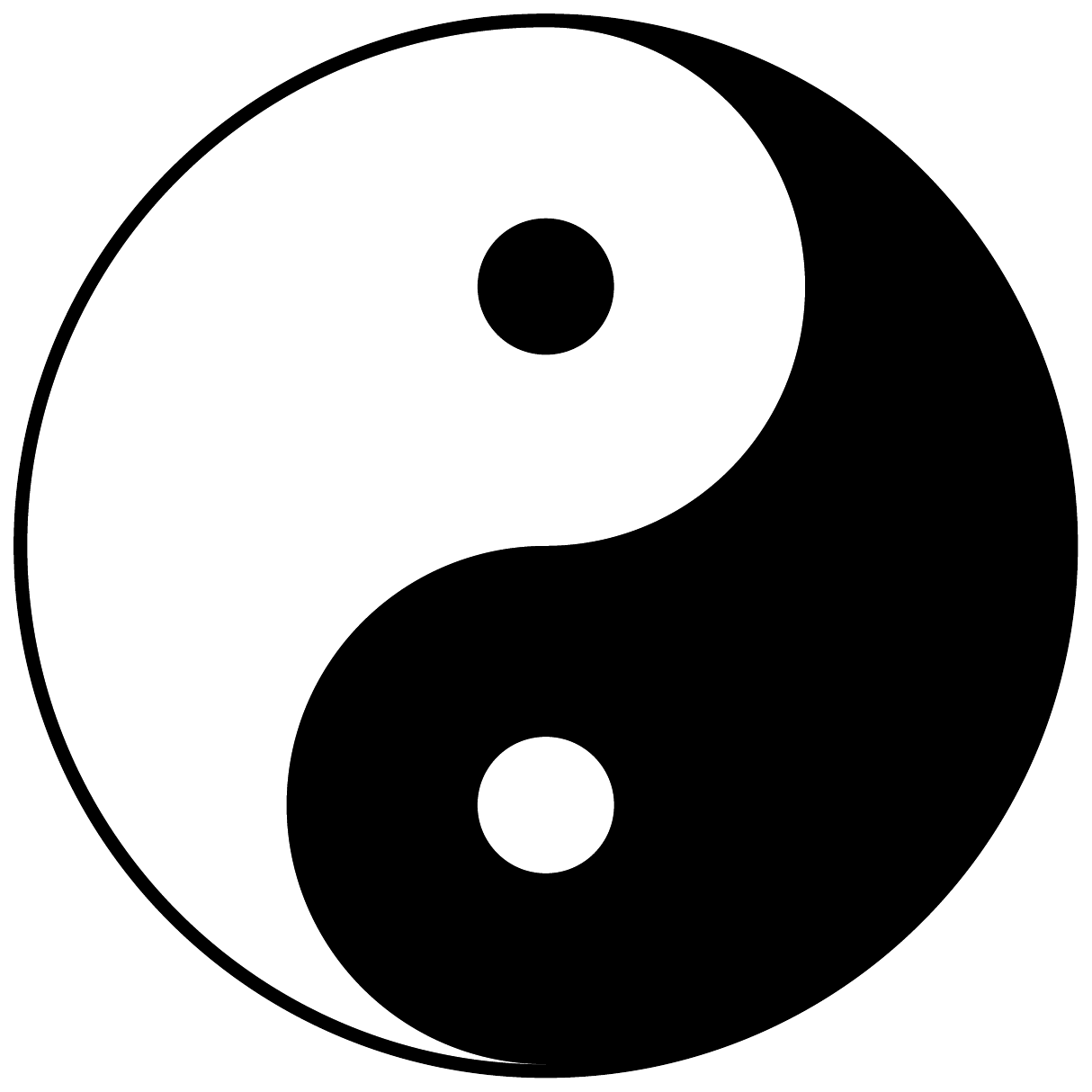}}

\def\BibTeX{{\rm B\kern-.05em{\sc i\kern-.025em b}\kern-.08emT\kern-.1667em\lower.7ex\hbox{E}\kern-.125emX}}
    
\copyrightyear{2019}
\acmYear{2019}
\acmConference[SC '19]{The International Conference for High Performance Computing, Networking, Storage, and Analysis}{November 17--22, 2019}{Denver, CO, USA}
\acmBooktitle{The International Conference for High Performance Computing, Networking, Storage, and Analysis (SC '19), November 17--22, 2019, Denver, CO, USA}
\acmPrice{15.00}
\acmDOI{10.1145/3295500.3357156}
\acmISBN{978-1-4503-6229-0/19/11}

\hypersetup{final}

\setcopyright{none}

\begin{document}

\newcommand{\todo}[1]{{\color{red} #1}}

\title[A Data-Centric Approach to Quantum Transport Simulations]{A Data-Centric Approach to Extreme-Scale \textit{Ab initio} Dissipative Quantum Transport Simulations}

\author{Alexandros Nikolaos Ziogas$^{*}$, Tal Ben-Nun$^{*}$, Guillermo Indalecio Fern\'{a}ndez$^{\scriptscriptstyle\dagger}$,
  Timo Schneider$^{*}$, Mathieu Luisier$^{\scriptscriptstyle\dagger}$, and Torsten Hoefler$^{*}$} 
\affiliation{%
$^*$Scalable Parallel Computing Laboratory, ETH Zurich, Switzerland\\
$^{\dagger}$Integrated Systems Laboratory, ETH Zurich, Switzerland 
}

\renewcommand{\shortauthors}{Ziogas et al.}

\begin{abstract}
The computational efficiency of a state of the art \textit{ab initio}
quantum transport (QT) solver, capable of revealing the coupled
electro-thermal properties of atomically-resolved nano-transistors,
has been improved by up to two orders of magnitude through a data
centric reorganization of the application. The approach yields
coarse-and fine-grained data-movement characteristics that can be used
for performance and communication modeling, communication-avoidance,
and dataflow transformations. The resulting code has been tuned
for two top-6 hybrid supercomputers, reaching a sustained performance
of 85.45 Pflop/s on 4,560 nodes of Summit (42.55\% of the peak)
in double precision, and 90.89 Pflop/s in mixed precision.
These computational achievements enable the restructured QT
simulator to treat realistic nanoelectronic devices made of more than
10,000 atoms within a 14$\times$ shorter duration than the original
code needs to handle a system with 1,000 atoms, on the same number of
CPUs/GPUs and with the same physical accuracy.
\end{abstract}


\maketitle


\section{Justification for Prize}
Record \textit{ab initio} dissipative quantum transport
simulation in devices made of $\geq$10,000 atoms (10$\times$
improvement w.r.t. state-of-the-art). Double precision
performance of 85.45 Pflop/s on 4,560 nodes of Summit (27,360
GPUs), and mixed precision of 90.89 Pflop/s. Reduction in time-to-solution per atom and communication volume 
by a factor of up to 140 and 136, respectively.


\vfill\eject

\section{Performance Attributes}

\begin{table}[h]
	\small
\begin{tabular}{l p{4cm}} 
\toprule
Performance attribute & Our submission \\
\midrule
Category of achievement & Scalability, time-to-solution\\
Type of method used & Non-linear system of equations\\
Results reported on basis of & Whole application including I/O\\
Precision reported & Double precision, Mixed precision\\
System scale & Measured on full-scale\\
Measurements & Timers, FLOP count, performance modeling\\
\bottomrule
\end{tabular}
\end{table}

\FloatBarrier

\begin{figure}
\centering
\includegraphics[width=.95\linewidth]{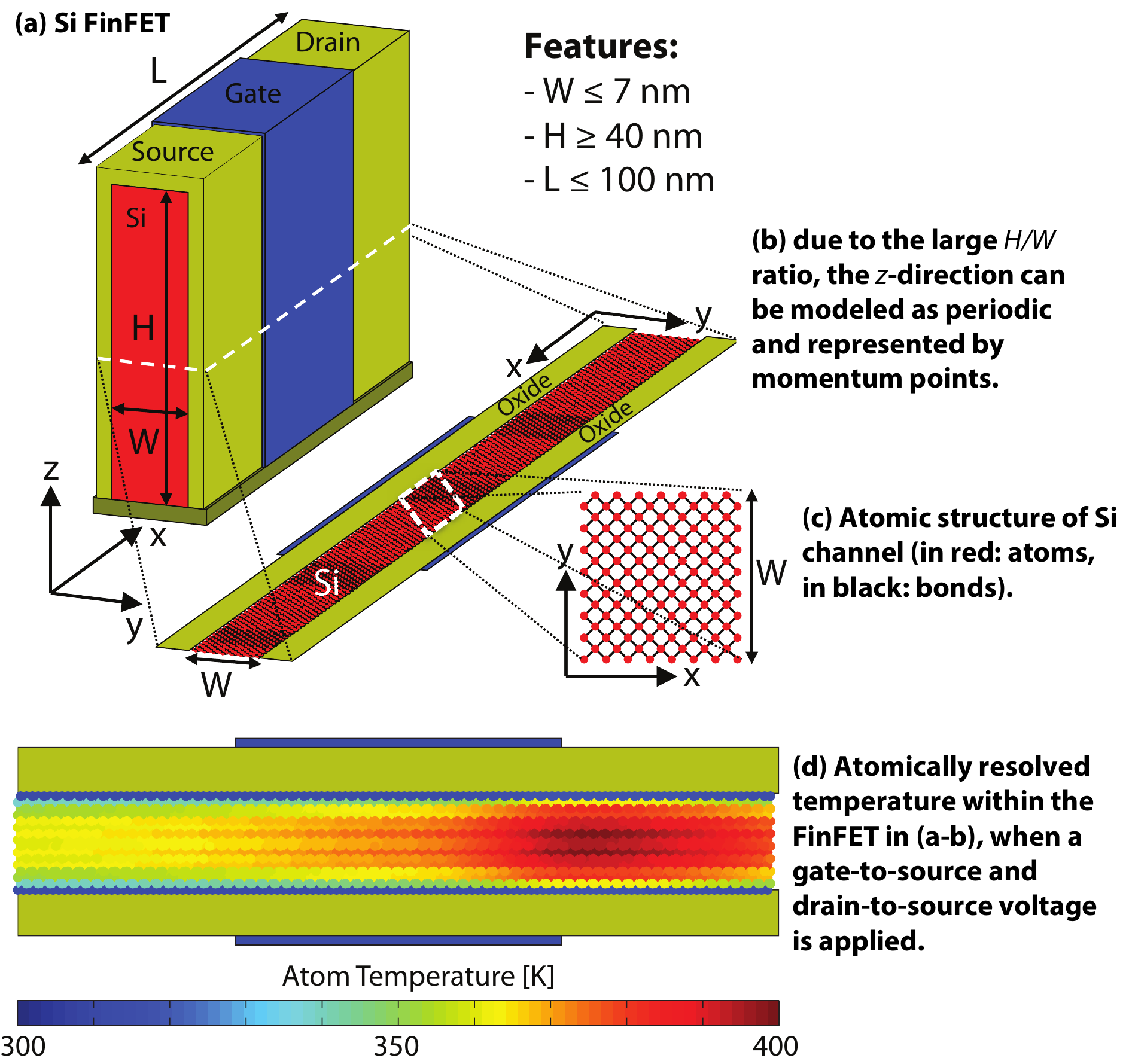} 
\vspace{-1em}
\caption{Self-heating effect simulation in a Silicon FinFET.}
\vspace{-2em}
\label{fig:res}
\end{figure}

\section{Overview of the Problem}\label{overview}

Much of human technological development and scientific advances in
the last decades have been driven by increases in computational
power. The cost-neutral growth provided by Moore's scaling law is however
coming to an end, threatening to stall progress in
strategic fields ranging from scientific simulations and bioinformatics
to machine learning and the Internet-of-Things. 
All these areas share an insatiable need for compute
power, which in turn drives much of their innovation. 
Future improvement of these fields, and science in general, will depend on 
overcoming fundamental technology barriers, one of which is the 
soon-to-become unmanageable heat dissipation in compute units.
\clearpage

Heat dissipation in microchips reached alarming peak values of 100
W/cm$^2$ already in
2006~\cite{wei,pop}. This led to the end of Dennard scaling and the
beginning of the ``multicore crisis'', an era with energy-efficient
parallel, but sequentially slower multicore CPUs.
Now, more than ten years later, average power densities of up to 30
W/cm$^2$, about four times more than hot plates, are commonplace in
modern high-performance CPUs, putting thermal management at the center
of attention of circuit designers~\cite{hotplate}. 
By scaling the dimensions of transistors more rapidly than their supply
voltage, the semiconductor industry has kept increasing heat dissipation
from one generation of microprocessors to the other. 
In this context, large-scale data and supercomputing centers are facing
critical challenges regarding the design and cost of their cooling
infrastructures. 
The price to pay for that has become exorbitant, as the cooling can consume up
to 40\% of the total electricity in data centers; a cumulative
cost of many billion dollars per year.

Landauer's theoretical limit of energy consumption for non-reversible
computing offers a glimmer of hope: today's processing units require
orders of magnitude more energy than the $k_BT\ln 2$ Joule bound to
(irreversibly) change one single bit. However, to approach this limit,
it will be necessary to first properly understand the mechanisms
behind nanoscale heat dissipation in semiconductor
devices~\cite{pop}. Fin field-effect transistors (FinFETs), as
schematized in Fig.~\ref{fig:res}(a-c), build the core of all recent
integrated circuits (ICs). Their dimensions do not 
exceed 100 nanometers along all directions, even 10 nm along one of
them (width $W$), with an active region composed of fewer than 1
million atoms. This makes them subject to strong quantum mechanical
and peculiar thermal effects. 

When a voltage $V_{ds}$ is applied across FinFETs, electrons start to
flow from the source to the drain contact, giving rise to an electrical
current whose magnitude depends on the gate bias $V_{gs}$. The potential
difference between source and drain allows electrons to transfer part
of their energy to the crystal lattice surrounding them. This energy
is converted into atomic vibrations, called phonons, that can propagate
throughout FinFETs. The more atoms vibrate, the ``hotter'' a device
becomes. This phenomenon, known as self- or Joule-heating, plays a
detrimental role in today's transistor technologies and has
consequences up to the system level. It is illustrated in
Fig.~\ref{fig:res}(d) (\S~\ref{sec:qt_sim} for details about
this simulation): a strong increase of the lattice temperature 
can be observed close to the drain contact of the simulated
FinFET. The negative influence of self-heating on CPU/GPU
performance can be minimized by devising computer-assisted strategies
to efficiently evacuate the generated heat from the active region of
transistors.

\subsection{Physical Model}\label{sec:problem}

Due to the large height/width ratio of FinFETs, heat generation and 
dissipation can be 
physically captured in a two-dimensional simulation domain 
comprising 10-15 thousand atoms and corresponding to a slice in the
$x$-$y$ plane. The height, aligned with the $z$-axis (see
Fig.~\ref{fig:res}(a-b)), can be treated as a periodic dimension and
represented by a momentum vector $k_z$ or $q_z$ in the range $[-\pi,
  \pi]$. The tiny width ($W\leq$7 nm) and length $L\leq$100 nm) of such
FinFETs require atomistic \textit{Quantum Transport (QT) simulation}
to accurately model and analyze their electro-thermal properties. In
this framework, electron and phonon (thermal) currents as well as
their interactions are evaluated by taking quantum mechanics into
account.  

The Non-equilibrium Green's Function (NEGF) formalism \cite{datta}
combined with density functional theory (DFT) \cite{kohn} lends itself
optimally to this type of calculations including electron and phonon
transport and thus to the investigation of self-heating in arbitrary
device geometries. With the help of DFT, an \textit{ab initio} method,
any material (combination) can be handled at the atomic level without
the need for empirical parameters.

\begin{figure}
	\centering
	\includegraphics[width=.8\linewidth]{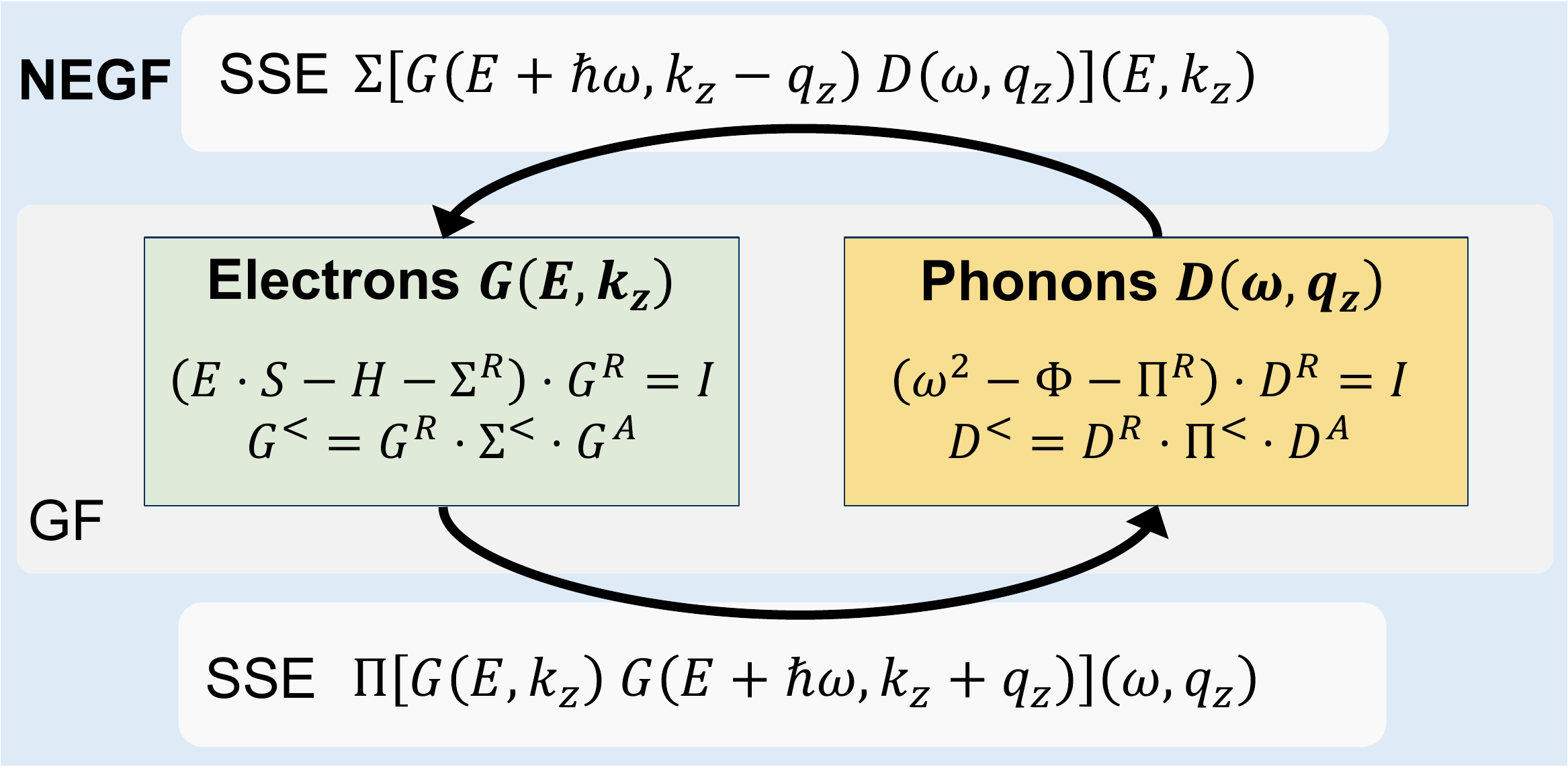} 
	\vspace{-1em}
	\caption{Self-consistent coupling between the GF and SSE phases (kernels) as
		part of the NEGF formalism.}
	\vspace{-1em}
	\label{fig:gf_sse}
\end{figure}

The DFT+NEGF equations for electron and phonon transport take the
form of a non-linear system of equations, as depicted in
Fig.~\ref{fig:gf_sse}. The electron ($G(E,k_z)$) and phonon
($D(\omega,q_z)$) Green's Functions (GF) at energy $E$, momentum
$k_z$/$q_z$, and frequency $\omega$ are coupled to each other
through scattering self-energies (SSE) $\Sigma(E,k_z)$ and
$\Pi(\omega,q_z)$ that depend on $[G(E\pm\hbar\omega,k_z-q_z)$
  $D(\omega,q_z)]$ and $[G(E,k_z)$ $G(E+\hbar\omega,k_z+q_z)]$, 
respectively. 

The electron and phonon Green's Functions are solved for all
possible electron energy ($N_E$) and momentum ($N_{k_z}$) points as
well as all phonon frequencies ($N_{\omega}$) and momentum
($N_{q_z}$). In case of self-heating, the difficulty does not come
from the solution of the GF equations, which are independent from
each other and have received a wide attention before \cite{sc11,sc15},
but from the fact that the scattering self-energies $\Sigma$ and $\Pi$
connect different energy-momentum ($E$, $k_z$) and frequency-momentum
($\omega$, $q_z$) pairs together. 

To obtain the electrical and energy currents or the temperature
distribution (see Fig.~\ref{fig:res}(d)) of a given device, the
non-linear GF and SSE equations must be iteratively solved until
convergence is reached. Depending on the geometry and bias conditions,
between $N_{iter}$=20 and 100 iterations are needed for that. The
algorithm starts by setting $\Sigma(E,k_z)$=$\Pi(\omega,q_z)$=0 
and continues by computing all GFs under this condition. The latter
then serve as inputs to the next phase, where the SSE are evaluated
for all ($E$, $k_z$) and ($\omega$, $q_z$).

\begin{table*}
	\caption{State of the Art Quantum Transport Simulators}
	\small
	\vspace{-1em}
	\begin{tabular}{l lll lll rc} 
		\toprule
		\textbf{Name} &	\multicolumn{6}{c}{\bf Maximum \# of Computed Atoms} & \multicolumn{2}{c}{\bf Scalability} \\	\cmidrule(l{2pt}r{2pt}){2-7}
		&\multicolumn{3}{c}{\bf Tight-binding-like$^*$} & \multicolumn{3}{c}{\bf DFT} & Max. Cores & Using\\\cmidrule(l{2pt}r{2pt}){2-4}\cmidrule(l{2pt}r{2pt}){5-7} 
		& $GF^\dagger_e$ & $GF^\dagger_{ph}$ & $GF+SSE$ & $GF^\dagger_e$ & $GF^\dagger_{ph}$ & $GF+SSE$ & (Magnitude) & GPUs \\
		\midrule
		GOLLUM~\cite{ferrer2014gollum}&1k&1k&---&100&100&---&N/A&\nocheck\\
		Kwant~\cite{groth2014kwant}&10k&---&---&---&---&---&N/A&\nocheck\\
		NanoTCAD ViDES~\cite{vides}&10k&---&---&---&---&---&N/A&\nocheck\\
		QuantumATK~\cite{atk}&10k&10k&---&1k&1k&---&1k&\nocheck\\		
		TB\_sim~\cite{niquet}&100k&---&10k$^\ddagger$&1k&---&---&10k&\yescheck\\		
		NEMO5~\cite{nemo5}&100k&100k&10k$^\ddagger$&---&---&---&100k&\yescheck \\
		OMEN~\cite{omen}&100k (1.44 Pflop/s \cite{sc11}) &100k&10k&10k  (15 Pflop/s \cite{sc15})&10k&1k (0.16 Pflop/s)&100k&\yescheck\\		
		\midrule
		\textbf{This work} & N/A & N/A & N/A & 10k & 10k & \textbf{10k (85.45 Pflop/s)} & \textbf{1M} & \yescheck \\
		\bottomrule
	\end{tabular}\\
	{\footnotesize $^*$: including Maximally-Localized Wannier Functions (MLWF), $\dagger$: Ballistic, $\ddagger$: Simplified.}\vspace{-1em}
	\label{tab:competitors}
\end{table*}

\subsection{Computational Challenges}

An intuitive algorithm to practically solve the GF+SSE system 
on supercomputers 
consists of
two loops: one over the momentum points ($k_z$/$q_z$) and another one
over the electron energies ($E$). This loop schedule results in
complex execution dependencies and communication patterns. 
The communication overhead quickly becomes a bottleneck with increasing number of
atoms and computational resources, crossing the petabyte range per iteration
for realistic transistor sizes (\S~\ref{sec:commmodel}).
Therefore, current simulations are limited to the order of 1,000 atoms, a value
much below what is needed to apply QT simulations in practical applications. 

Even in a scenario where each node operates at maximum computational
efficiency, large-scale QT simulations are bound by both communication
volume and memory requirements.
The former inhibits strong scaling, as simulation time includes
nanostructure-dependent point-to-point communication patterns, which
become infeasible when increasing node count. 
The memory bottleneck is a direct result of the former. It hinders
large simulations due to the increased memory requirements w.r.t. atom count. 
Transforming the QT simulation algorithm to minimize communication is
thus the key to simultaneously model larger devices and increase
scalability on different supercomputers.

\section{Current State of the Art}

Quantum transport simulation is an important driver of innovation in
nanoelectronics. Thus, many atomistic quantum transport simulators
that can model the characteristics of nano-devices have been
developed~\cite{ferrer2014gollum,groth2014kwant,nemo5,atk,vides,niquet,omen}.
Their performance is summarized in Table \ref{tab:competitors}, where their
estimated maximum number of atoms that can be simulated for a given physical
model is provided.
Only orders of magnitude are shown, as these quantities depend on the device
geometries and band-structure method.
It should be mentioned that most tools are limited to tight-binding-like (TB)
Hamiltonians, because they are computationally cheaper than DFT ones
(less orbitals and neighbors per atom).
This explains the larger systems that can be treated with TB.
However, such approaches are \emph{not accurate enough} when it comes to
the exploration of material stacks, amorphous layers, metallic contacts,
or interfaces as needed in transistor design.
In these cases, the \emph{higher accuracy of DFT is required} and leads to a much higher
computational cost.

When it comes to the modeling of self-heating at the \textit{ab
  initio} level, the following NEGF equations must be solved for
electrons:
\begin{eqnarray}
\left\{
\begin{array}{l}
\left(E\cdot{\mathbf S}(k_z)-{\mathbf H}(k_z)-{\mathbf \Sigma}^{R}(E,k_z)\right)\cdot 
{\mathbf G}^{R}(E,k_z)={\mathbf I}\\
{\mathbf G}^{\gtrless}(E,k_z)=\mathbf{G}^{R}(E,k_z)\cdot{\mathbf
  \Sigma}^{\gtrless}(E,k_z)\cdot{\mathbf G}^{A}(E,k_z).
\end{array}
\right.
\label{eq:1}
\end{eqnarray}
In Eq.~(\ref{eq:1}), ${\mathbf S}(k_z)$ and ${\mathbf H}(k_z)$ are the
$k_z$-dependent overlap and Hamiltonian matrices, respectively. They
must be produced by a DFT code with a localized basis set (here: CP2K
\cite{cp2k}) and have a size $N_a\times N_{orb}$ ($N_a$: total number
of atoms, $N_{orb}$: number of orbitals per atom). The ${\mathbf
  G}(E,k_z)$ electron GFs have the same size as ${\mathbf S/H}$
and ${\mathbf I}$, the identity matrix. They can be either retarded
($R$), advanced ($A$), lesser ($<$), or greater ($>$). The same
notation applies to the self-energies ${\mathbf\Sigma}(E,k_z)$ that
include a boundary and scattering (superscript $S$) term.

To handle phonon transport, a similar GF system of equations
must be processed: the electron energy $E$ is replaced by the square
of the phonon frequency $\omega^2$, the Hamiltonian $H(k_z)$ by the
dynamical matrix ${\mathbf \Phi}(q_z)$, and $S(k_z)$ by the identity 
matrix $I$ with $N_{orb}$=$N_{3D}$=3, the three axes along which
crystals can vibrate.

Eq.~(\ref{eq:1}) and its phonon equivalent can be solved with a
so-called recursive Green's Function (RGF) algorithm \cite{rgf}. All
matrices ($H$, $S$, and $\Phi$) are block-tri-diagonal and can be
divided into $bnum$ blocks with $\frac{N_a}{bnum}$ atoms each, if the
structure is homogeneous, as here. RGF then performs a
forward/backward pass over the $bnum$ blocks. It has been demonstrated
that \textit{converting selected operations of RGF to single-precision
typically leads to inaccurate results} \cite{sc11}.

The electron ($\Sigma^S$) and phonon ($\Pi^S$) scattering self-energies 
(less-er and greater components) can be written as follows \cite{stieger}: 

{\small
\begin{eqnarray}
{\mathbf \Sigma}^{\gtrless S}_{aa}(E,k_z)=i\sum_{q_zijl}\int
\frac{d\hbar\omega}{2\pi}\left[\nabla_i{\mathbf H}_{ab}\cdot 
{\mathbf G}^{\gtrless}_{bb}(E-\hbar\omega,k_z-q_z)\cdot\right.\nonumber\\
\nabla_j{\mathbf H}_{ba}\cdot\left({\mathbf D}^{\gtrless ij}_{ba}(\omega,q_z)-
{\mathbf D}^{\gtrless ij}_{bb}(\omega,q_z)-\right.\nonumber\\
\left.\left.{\mathbf D}^{\gtrless ij}_{aa}(\omega,q_z)+{\mathbf D}^{\gtrless
  ij}_{ab}(\omega,q_z)\right)\right],\label{eq:3}
\end{eqnarray}
}
{\small
\begin{eqnarray}
{\mathbf \Pi}^{\gtrless S}_{ab}(\omega,q_z)=-i\sum_{k_zl}\int\frac{dE}{2\pi} 
\mathrm{tr}\left\{\nabla_{i}{\mathbf H}_{la}\cdot{\mathbf G}^{\gtrless}_{aa}(E+\hbar\omega,k_z+q_z)\cdot\right.\nonumber\\
&\hspace*{-6cm}\left.\nabla_j{\mathbf H}_{al}\cdot{\mathbf G}^{\lessgtr}_{ll}(E,k_z)\right\}.
\label{eq:4}
\end{eqnarray}
}
In Eq.~(\ref{eq:4}), the sum over $l$ is replaced by $l$=$b$, if
$a$=$b$. All Green's Functions ${\mathbf G_{ab}}$ (${\mathbf D_{ab}}$)
are matrices of size $N_{orb}\times N_{orb}$ ($N_{3D}\times
N_{3D}$). They describe the coupling between two neighbor atoms $a$ and
  $b$ at positions ${\mathbf R}_a$ and 
  ${\mathbf R}_b$. Each atom has $N_b$ neighbors. The term
$\nabla_i {\mathbf H}_{ab}$ is the derivative of the $\mathbf{H}_{ab}$
Hamiltonian coupling atoms $a$ and $b$. It is computed
  with DFT. Only the diagonal blocks of ${\mathbf
    \Sigma}^{R\gtrless,S}$ and $N_b$ non-diagonal blocks of ${\mathbf
    \Pi}^{R\gtrless,S}$ are considered.

The evaluation of Eqs.~(\ref{eq:3}-\ref{eq:4}) does not require the
knowledge of all entries of the ${\mathbf G}$ and ${\mathbf D}$
matrices, but of two 5-D tensors of shape
$[N_{k_z}, N_E, N_a, N_{orb}, N_{orb}]$ for electrons and two 6-D
tensors of shape $[N_{q_z}, N_{\omega}, N_a, N_b + 1, N_{3D},
  N_{3D}]$ for phonons. Each $[k_z, E, N_a, N_{orb}, N_{orb}]$ and
$[q_z, \omega, N_a, N_b + 1, N_{3D},$ $N_{3D}]$ combination is
produced independently from the other by solving the GF
equations with RGF. The electron and phonon SSE can also be reshaped
into multi-dimensional tensors with the same dimensions as their GF
counterparts, but they cannot be computed independently due to energy,
frequency, and momentum coupling. 

To the best of our knowledge, the only tool that can solve
Eqs.~(\ref{eq:1}) to (\ref{eq:4}) self-consistently, in structures
composed of thousands of atoms, at the DFT level is OMEN \cite{omen},
a \textit{\textbf{two-time Gordon Bell Prize finalist}}
\cite{sc11,sc15}.\footnote{Previous achievements: development 
of parallel algorithms to deal with ballistic transport
(Eq.~(\ref{eq:1}) alone) expressed in a tight-binding (SC11) or DFT
(SC15) basis.} The application is written in C++, contains 90,000 lines of
code in total, and uses MPI as its communication protocol. Some parts
of it have been ported to GPUs using the CUDA language and take
advantage of libraries such as cuBLAS, cuSPARSE, and MAGMA. The
electron-phonon scattering model was first implemented based on the
tight-binding method and a three-level MPI distribution of the
workload (momentum, energy, and spatial domain decomposition). A first
release of the model with equilibrium phonon (${\mathbf \Pi^S}$=0) was
validated up to 95k cores for a device with $N_a$=5,402, $N_b$=4,
$N_{orb}$=10, $N_{k_z}$=21, and $N_E$=1,130. These runs showed that
the application can reach a parallel efficiency of 57\%, when going
from 3,276 up to 95,256 cores, with the SSE phase consuming from 25\%
to 50\% of the total simulation times. The reason for the increase in SSE time
could be attributed to the communication time required to gather all
Green's Function inputs for Eq.~(\ref{eq:3}), which grew from 16 to
48\% of the total simulation time~\cite{sc10} as the number of cores
went from 3,276 to 95,256. 

After extending the electron-phonon scattering model to DFT and adding
phonon transport to it, it has been observed that the time spent in
the SSE phase (communication and computation) explodes. Even for a
small structure with $N_a$=2,112, $N_{orb}$=4, $N_{k_z}$=$N_{q_z}$=11,
$N_E$=650, $N_{\omega}$=30, and $N_b$=13, 95\% of the total simulation
time is dedicated to SSE, regardless of the number of used
cores/nodes, among which $\sim$60\% for the communication between
the different MPI tasks. The relevance of this model is therefore limited.

\begin{figure*}
	\begin{minipage}[b]{.32\linewidth}
		\captionsetup{type=table}
		\centering
		\footnotesize
		\caption{Requirements for Accurate\\ Dissipative DFT+NEGF
			Simulations}\vspace{-1em}
		\begin{tabular}{l p{2.75cm} r} 
			\toprule
			\textbf{Variable} & \textbf{Description} & \textbf{Value} \\
			\midrule
			$N_{k_z}/N_{q_z}$ & Number of electron/phonon
			momentum points & $\geq$21 \\			
			$N_E$ & Number of energy points & $\geq$1,000 \\
			$N_\omega$ & Number of phonon frequencies	& $\geq$50 \\
			$N_a$ & Total number of atoms per device
			structure & $\geq$10,000\\
			$N_b$ & Neighbors considered for each atom
			& $\geq$30 \\
			$N_{orb}$ & Number of orbitals per atom & $\geq$10 \\
			$N_{3D}$ & Degrees of freedom for crystal vibrations & 3 \\
			\bottomrule
		\end{tabular}
		\vspace{3.5em}
		\label{tab:common-values}
	\end{minipage}\hfill
	\begin{minipage}[b]{.664\linewidth}
		\centering
		\includegraphics[width=.95\linewidth, page=14]{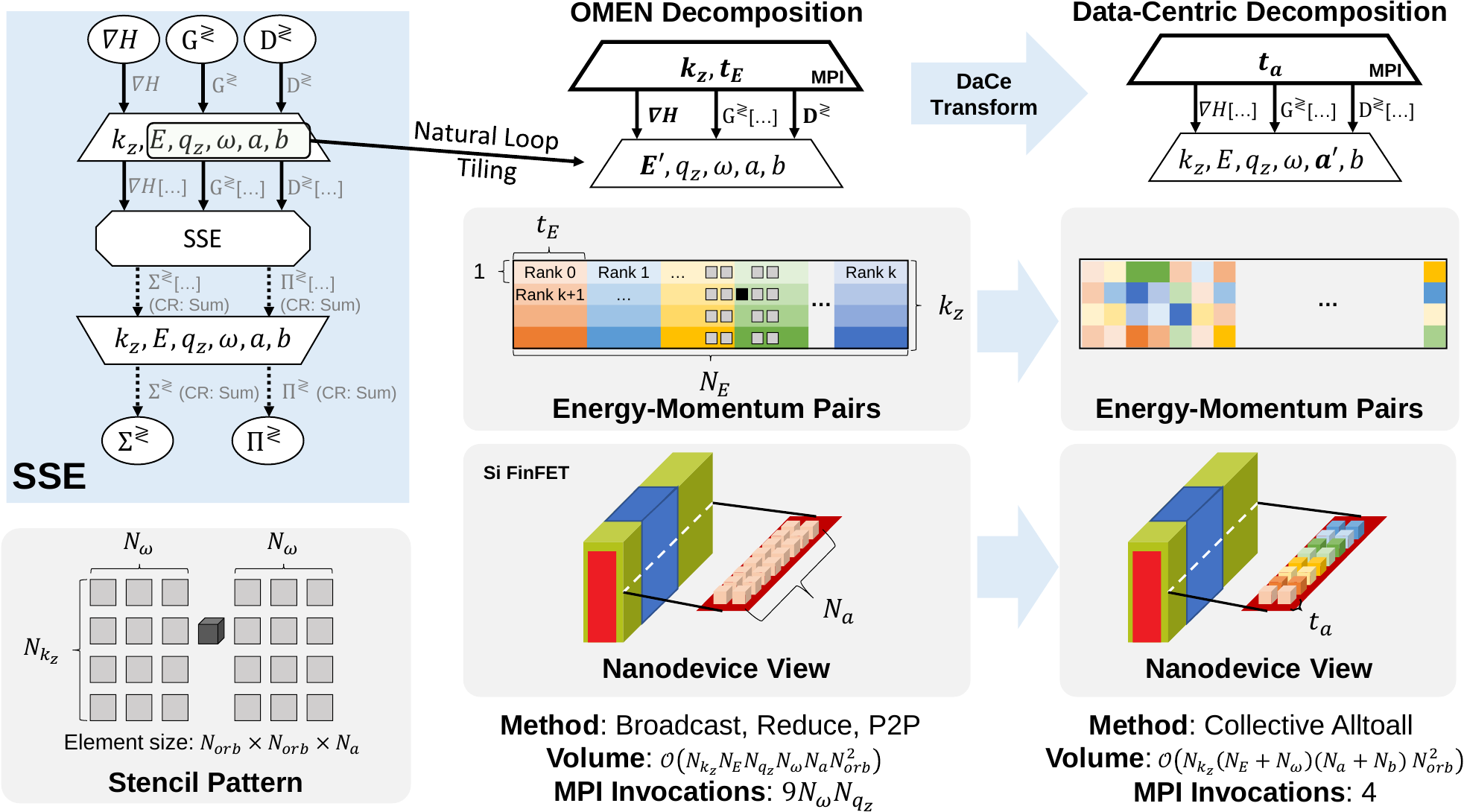}
		\vspace{-1em}
		\caption{Domain decomposition of SSE in OMEN and DaCe.}
		\vspace{-1em}
		\label{fig:ssetiling}
	\end{minipage}
\end{figure*}

\textbf{Understanding realistic FinFET transistors requires simulations
with $\mathbf{N_a\geq 10,000}$ atoms and high accuracy
($\mathbf{N_{k_Z}}$$\mathbf{>}$$\mathbf{20}$, $\mathbf{N_E}$$\mathbf{>}$$\mathbf{1,000}$}, see Table \ref{tab:common-values}). 
In order to achieve a reasonable cost (\textit{time, money, and energy}), the
algorithms to solve Eqs.~(\ref{eq:1}) to (\ref{eq:4}) must be
drastically improved.
\textbf{The required algorithmic improvements needed are at least one order of
magnitude in the number of atoms and two orders of magnitude in
computational time per atom.}
In the following sections, we demonstrate how both can be achieved with a
novel data-centric view of the simulation. 

\begin{figure}[t]
	\centering
	\includegraphics[width=\linewidth,page=3]{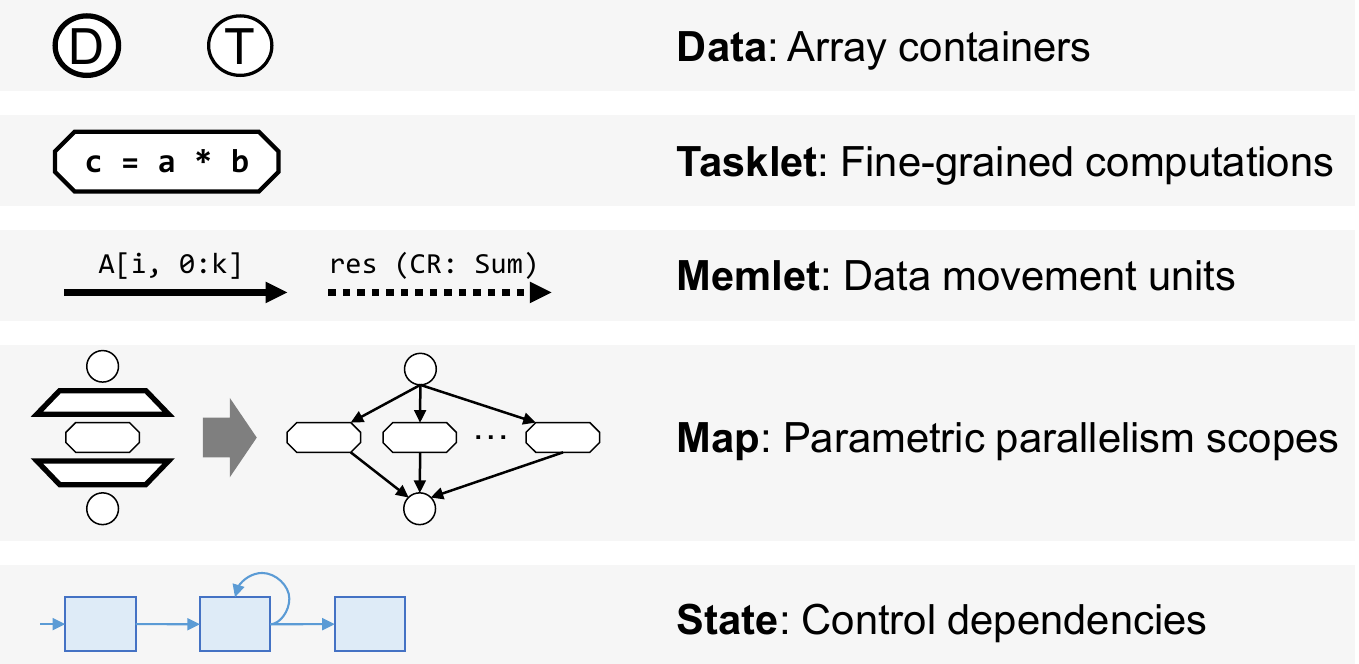}
	\vspace{-2em}
	\caption{Stateful Dataflow Multigraph (SDFG) concepts.}
	\vspace{-2em}
	\label{fig:sdfg}
\end{figure}

\section{Innovations Realized}

The discussed self-heating effects have formed the landscape of current
HPC systems, which consist of new architectures attempting to work around the
physical constraints. Thus, no two cluster systems are the same and
heterogeneity is commonplace.
Each setup requires careful tuning of application performance, focused mostly 
on data movement, which causes the lion's share of energy dissipation~\cite{padal}. 
As this kind of tuning demands in-depth knowledge of the hardware, it
is typically performed by a \yang~\textit{Performance Engineer}, a
developer who is versed in intricate system details, existing
high-performance libraries, and capable of modeling performance and
setting up optimized procedures independently.
This role, which complements the \yin~\textit{Domain Scientist}, has been 
increasingly important in scientific computing
for the past three decades, but is now essential for any application 
beyond straightforward linear algebra to operate at extreme scales. 
Until now, both Domain Scientists and Performance Engineers would work 
with one code-base. This creates a co-dependent
\yinyang~situation~\cite{pat-maccormicks-sos-talk}, where the original
domain code is tuned to a point that modifying the
algorithm or transforming its behavior is difficult to one without the
presence of the other, even if data locality or computational
semantics are not changed.  

\begin{figure}[t]
	\centering 
	\includegraphics[width=.86\linewidth, page=2]{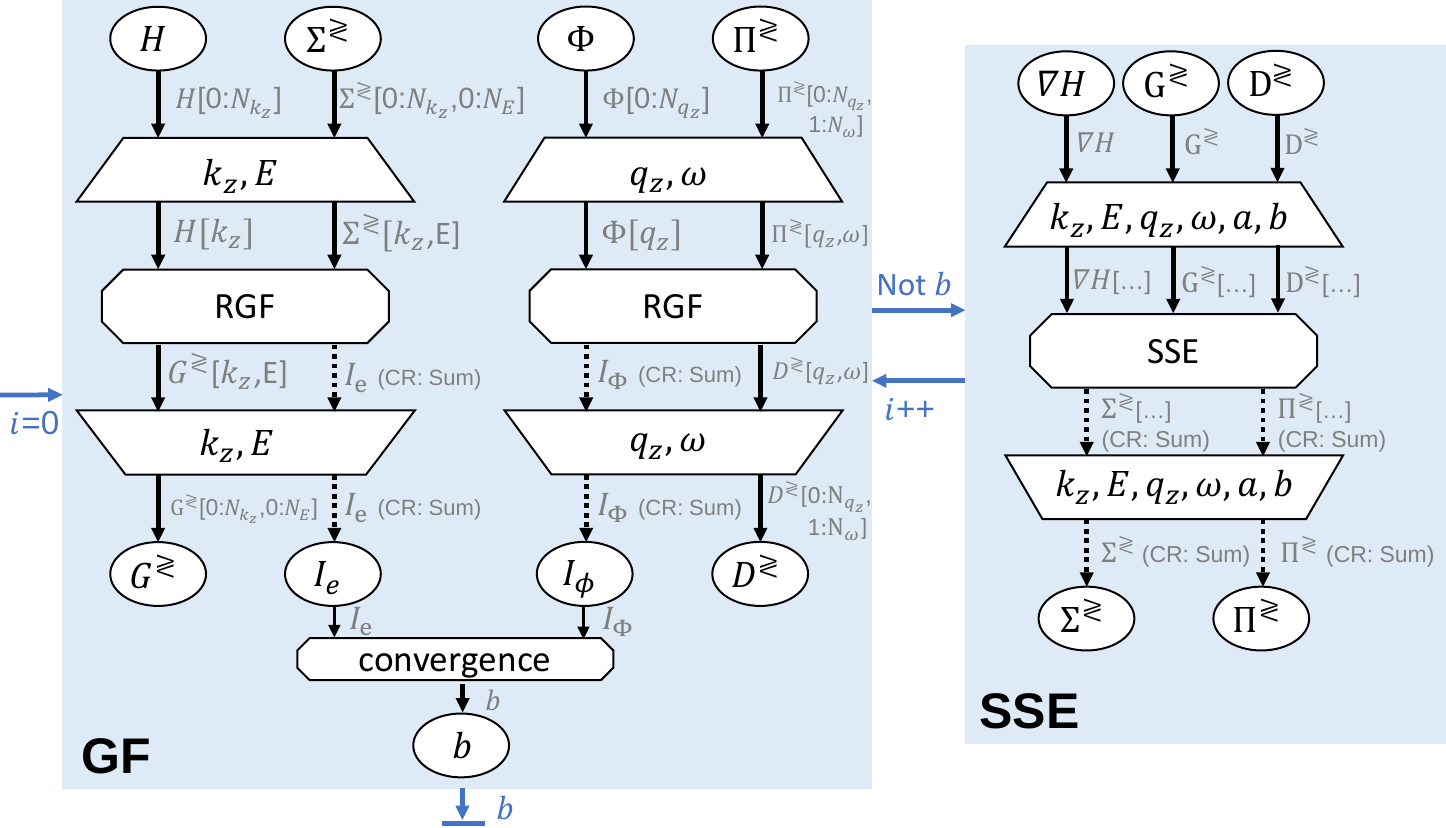}
	\vspace{-1em}
	\caption{SDFG of the entire simulation.}
	\vspace{-1em}
	\label{fig:dace-view}
\end{figure}

We propose a paradigm change by rewriting the quantum
transport problem as implemented in OMEN from a data-centric
perspective.  
We show that the key to eliminating the scaling bottleneck is 
in formulating a communication-avoiding algorithm, which
is tightly coupled with recovering local and global data dependencies
of the application. 
We start from a reference Python implementation, using Data-Centric
(DaCe) Parallel Programming~\cite{sdfg} to express the computations
separately from data movement (Fig.~\ref{fig:sdfg}). DaCe automatically constructs a
\textit{stateful dataflow} view (Fig.~\ref{fig:dace-view}) that can be used to optimize data
movement without modifying the original computation. This enables
both rethinking the communication pattern of the simulation, and tuning the
data movement for each target supercomputer. 
In the remainder of this paper, the new code is referred to as
DaCe OMEN or simply DaCe.

We report the following innovations, most of them being directly obtained as a result of the data-centric view:
\begin{itemize}
	\item \textbf{Data Ingestion}: We stage the material and use chunked broadcast to deliver it to nodes. This reduced Piz Daint start-up time at full-scale from $\sim$30 minutes to under two.
	\item \textbf{Load Balancing}: Similar to OMEN, we divide work among electrons and phonons unevenly, so as to reduce imbalance.
	\item \textbf{Communication Avoidance}: We reformulate communication
    in a non-natural way from a physics perspective, leading to two
    orders of magnitude reduction in volume.
	\item \textbf{Boundary Conditions}: We pipeline contour integral calculation on the GPUs, computing concurrently and accumulating resulting matrices using on-GPU reduction.
	\item \textbf{Sparsity Utilization}: We tune and investigate different data-centric transformations on sparse Hamiltonian blocks in GF, using a combination of sparse and dense matrices.
	\item \textbf{Pipelining}: The DaCe framework automatically generates copy/compute and compute/compute overlap, resulting in 60 auto-generated CUDA streams.
	\item \textbf{Computational Innovations}: We reformulate SSE computations using data-centric transformations. Using fission and data layout transformations, we reshape the job into a stencil-like strided-batched GEMM operation, where the DaCe implementation yields up to 4.8$\times$ speedup over cuBLAS.
\end{itemize}

The full implementation details and transformations are described by Ziogas et al.~\cite{ziogas19}. Below, we highlight the innovations that led to the most significant performance improvements.

\subsection{Data-Centric Parallel Programming}\label{sec:dace}

Communication-Avoiding (CA) algorithms~\cite{demmel13ca,writeav} are defined as algorithm variants and schedules 
(orders of operations) that minimize the total number of performed memory loads and stores,
achieving lower bounds in some cases. 
To achieve such bounds, a subset of those algorithms is \textit{matrix-free}\footnote{The term is derived from solvers that do not need to store the entire matrix in memory.}.
A key requirement in modifying an algorithm to achieve communication avoidance is to explicitly
formulate its data movement characteristics. 
The schedule can then be changed by reorganizing the data flow to minimize the sum of
accesses in the algorithm.
Recovering a data-centric view of an algorithm, which makes movement explicit 
throughout all levels (from a single core to the entire cluster), is thus the path 
forward in scaling up the creation of CA variants to more complex algorithms and multi-level memory hierarchies as one.

DaCe defines a development workflow where the original algorithm is independent from its
data movement representation, enabling symbolic analysis and transformation of the latter
without modifying the scientific code. 
This way, a CA variant can be formulated and developed by a performance engineer, while 
the original algorithm retains readability and maintainability.
At the core of the DaCe implementation is the Stateful DataFlow
multiGraph (SDFG) \cite{sdfg}, an intermediate representation that
encapsulates data movement and can be generated from high-level code in
Python.
The syntax (node and edge types) of SDFGs is listed in
Fig.~\ref{fig:sdfg}.

The workflow is as follows:
The domain scientist designs an algorithm and implements it with linear algebra operations (imposing dataflow implicitly), or with Memlets and Tasklets (specifying dataflow explicitly).
The Memlet edges define all data movement, which is seen in the input and output of each Tasklet, but also entering and leaving Maps with their overall requirements \textit{and} total number of accesses.
This implementation is then parsed into an SDFG, 
on which performance engineers may apply
graph transformations to improve data locality. 
After transformation, the optimized SDFG is compiled to machine code for performance 
evaluation. It may be further transformed interactively and tuned for different target platforms and memory hierarchy characteristics.
The SDFG representation allows the performance engineer to add local arrays, reshape and nest Maps (e.g., to impose a tiled schedule), fuse scopes, map computations to accelerators (GPUs, FPGAs), and other transformations that may modify the overall number of accesses.

\begin{figure*}
	\centering 
	\includegraphics[width=\linewidth]{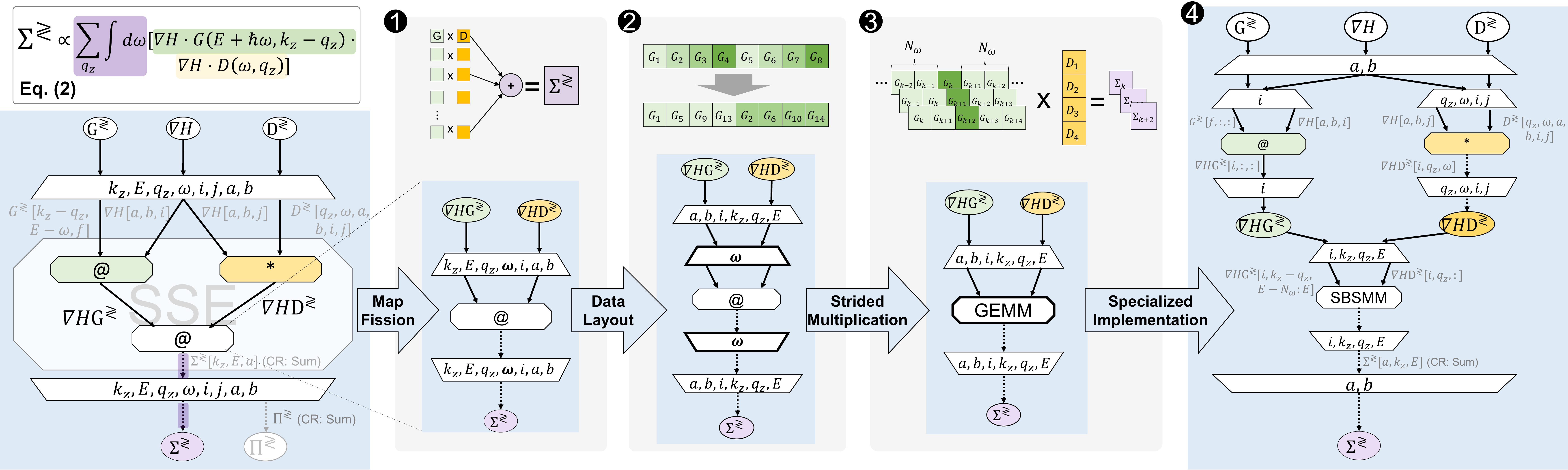}
	\vspace{-2em}
	\caption{Transforming the dataflow of SSE using SDFGs.}
	\vspace{-1em}
	\label{fig:transformation}
\end{figure*}

The top-level view of the simulation (Fig.~\ref{fig:dace-view}) shows that
it iterates over two states, GF and SSE.
The former computes the boundary conditions, cast them into self-energies, solve for the Green's Functions, and extract physical observables (current, density) from them. The state consists of two concurrent Maps, one for
the electrons and one for the phonons (\S~\ref{sec:problem}).
The SSE state computes the scattering self-energies ${\mathbf \Sigma}^\gtrless$ and ${\mathbf \Pi}^\gtrless$.
At this point, we opt to represent the RGF solvers and SSE kernel as Tasklets, i.e., collapsing their dataflow, so as to focus on high-level aspects of the algorithm. 
This view indicates that the RGF solver cannot compute the Green's Functions for a specific atom separately from the rest of the material (operating on all atoms for a specific energy-momentum pair), and that SSE 
outputs the contribution of a specific
$\left(k_z, E, q_z, \omega, a, b\right)$ point to ${\mathbf \Sigma}^\gtrless$ and ${\mathbf \Pi}^\gtrless$.
These contributions are then accumulated to the output tensors, as indicated by the dotted Memlet edges.
The accumulation is considered associative; therefore the map can compute all dimensions of the inputs and outputs in parallel.
Below we show how the data-centric view
is used to identify and implement a \textit{tensor-free} CA variant of OMEN,
achieving near optimal communication for the first time
in this scientific domain.

\vspace{-0.5em}
\subsection{Communication Avoidance}\label{sec:ca}

Figure ~\ref{fig:ssetiling} describes the domain decomposition of SSE
computation in OMEN and the DaCe variant, while relevant parameters are given in Table \ref{tab:common-values}. The main part of the
computation uses a complex stencil pattern (bottom left of figure) to
update 3-D tensors (black cube). In the 2-D stencil, neighboring tensors
(grey boxes) are multiplied and accumulated over one full dimension
($N_{k_z}$) and one with a radius of $N_\omega$ tensors on each side.
The domain scientists who first implemented OMEN naturally decomposed
the 6D loop nest along the first two dimensions into a
$k_z\times E/t_E$ process grid of energy-momentum pairs (middle
part of the figure). 
In the data-centric view, this original decomposition is expressed as a
tiling transformation of the SDFG, where the outermost (top) map
controls process mapping. Through sophisticated use of MPI communicators
(grouped by rank brightness in the figure) and collectives, OMEN can use
broadcast and reduction operations to distribute the data across nodes.

Upon inspecting data movement in the SDFG Memlets, this decomposition yields full data dependencies (Fig.~\ref{fig:ssetiling}, top) and a multiplicative expression for the number of accesses (bottom of figure). 
If the map is, however, tiled by the atom positions on the
nano-device instead (which Eq.~(\ref{eq:3}) does
not expose, as it computes one pair), much of the movement can be
reduced, as indicated in Fig.~\ref{fig:ssetiling}.

The resulting pair decomposition is rather complex, which would
traditionally take an entire code-base rewrite to support, but in our
case uses only two graph transformations on the SDFG. We make the
modification shown in the top-right of Fig.~\ref{fig:ssetiling}, which
leads to an asymptotic reduction in communication volume, speedup of two
orders of magnitude, and a reduction in MPI calls to a constant number
as a byproduct of the movement scheme (\S~\ref{sec:perfmodel}, \ref{sec:results}).

\vspace{-0.5em}
\subsection{Dataflow Optimizations}\label{sec:dataflow}

The data-centric view not only encompasses macro dataflow that imposes communication, but also data movement within compute devices. We use DaCe to transform all computations in the communication-avoiding variant of OMEN, including the RGF algorithm, SSE, and boundary 
conditions, and automatically generate GPU code. In Fig.~\ref{fig:transformation} we showcase a subset of these transformations, focusing on a
bottleneck subgraph of the simulator, which is found within the SSE kernel: computing 
${\mathbf \Sigma}^\gtrless$ as in Eq.~(\ref{eq:3}). We note that computation of ${\mathbf \Pi}^\gtrless$ is 
transformed in a similar manner.

An initial SDFG representation of the ${\mathbf \Sigma}^\gtrless$ computation is
shown on the left side of the figure.
In step \one, we apply Map Fission and isolate the individual operations to
separate Maps.
This allows storing intermediate results to transient arrays so that they can
be reused, effectively lowering the number of multiplications (akin to 
regrouping algebraic computations in the formula). We then transform each 
resulting map separately, but focus here on ${\mathbf \Sigma}^\gtrless$, 
which results from the
accumulation of numerous products of $N_{orb}^2$-sized matrices.
Since $N_{orb}$ is small (typically ranging between 10 and 25),
these multiplications are inefficient and must be transformed.

In step \two, we reorder the ${\mathbf \Sigma}^\gtrless$ map dimensions and apply data-layout
transformations on the input, transient and output arrays.
The new data layout enables representing the $\omega$ map as the innermost dimension and split it out, exposing linear-algebra optimizations.
In step \three, the individual multiplications are aggregated to
more efficient GEMM operations using the structure of the map.
Furthermore, due to the data re-layout, the inputs and outputs among
the sequential GEMM operations are accessed with constant stride.
This in turn allows us to use optimized strided-batched operations, such as
\texttt{cublasZgemmStridedBatched} from the cuBLAS library.
Finally, we specialize the strided-batched operation in DaCe, the performance of which is discussed in \S~\ref{sec:sbsmm}.

In step \four, the separate maps are fused back together.
The optimized SDFG representation of the ${\mathbf \Sigma}^\gtrless$
computation is depicted on the right-hand side SDFG.
All the transformations described above result in an optimized SSE kernel that
both reduces the flop count (\S~\ref{sec:compmodel})
and has increased computational efficiency (\S~\ref{sec:nodeperf}).

\vspace{-0.5em}
\subsection{Mixed-Precision Computation} \label{sec:mixedprec}

The iterative GF-SSE solver creates an opportunity to trade accuracy for performance. The computation of $D^\gtrless$ and $G^\gtrless$ in the RGF phase consists of deep data dependency graph, whose precision cannot be reduced without substantially impacting the result \cite{sc11}. However, the computation of $\Sigma^{\gtrless}$ in SSE, which is a sum of matrix multiplications, can benefit from half-precision and Tensor Cores.

We adapt the computation in Fig.~\ref{fig:transformation} to use NVIDIA GPU Tensor Cores by transforming the tensors to split-complex format (contiguous real followed by imaginary values), padding the internal matrices to the required 16$\times$16. For normalization, we observe that the dynamic range of the inputs for the multiplications depends on $\nabla H,D^\gtrless,G^\gtrless$, and compute factors based on their magnitudes. Algebraically, denormalization entails scaling by inverse factors. Out-of-range values are clamped to avoid under/overflow and minimize the difference over accumulation, done in double-precision.

\section{How Performance Was Measured}

\label{omen_struct}

To measure the performance of DaCe OMEN and compare it to the state of
the art (original code of OMEN), two Si FinFET-like structures similar
to the one shown in Fig.~\ref{fig:res} have been defined:
\begin{itemize}
\item The first one labelled ``Small'' is characterized by $W=$ 2.1nm
  and $L=$ 35nm. It exhibits the following parameters: $N_a=4{,}864,
  N_b=34, N_E=706$, $N_{\omega}=70$, and $N_{k_z}$/$N_{qz}$ varying
  between 3 and 11. Unrealistically small $N_E$ and
  $N_{k_z}$/$N_{q_z}$ have been chosen to allow the original
  version of OMEN to simulate it too, but the physical accuracy is 
  not ensured.
\item The second one labelled ``Large'' relies on realistic dimensions 
  ($W=$ 4.8nm and $L=$ 35nm) and physical parameters: $N_a=10{,}240,
  N_b=34, N_E=1{,}220$, $N_{\omega}=70$, and $N_{k_z}$/$N_{qz}$ varying
  from 5 to 21 to cover both weak and strong scaling.  
\end{itemize}

\subsection{Performance Model}\label{sec:perfmodel}

The majority of computations revolves around three kernels (\S\ref{sec:dace}): (a)
computation of the boundary conditions; (b)
Recursive Green's Function (RGF); and (c) the SSE kernel. 
The first two kernels represent most of the computational load of the
GF phase, while the SSE phase comprises the SSE kernel. 
Table \ref{tab:flop} shows the
flop values, defined empirically and analytically, for the ``Small'' Silicon
structure with
varying $N_{k_z}$ values.

\subsubsection{Computation Model}\label{sec:compmodel}
The kernels of the GF phase mainly involve matrix multiplications between both
dense and sparse matrices.
The computational complexity of the RGF algorithm for each $\left(k_z, E\right)$ point is
$8\cdot\left(26\cdot bnum-25\right)\left(\frac{N_aN_{orb}}{bnum}\right)^3 +
\mathcal{O}\left(\left(\frac{N_aN_{orb}}{bnum}\right)^3\right)$,\\
where $bnum$ is the number of diagonal Hamiltonian blocks.
The first term accounts for the dense operations, which comprise 90\% of the flop count.
The latter term is an upper bound on the computational load of the sparse operations.
Since the RGF kernel is executed on the GPU, we measure the exact GPU flop count with the NVIDIA profiler \texttt{nvprof}.

In the SSE phase, the
flop count of each dense small matrix
multiplication (sized $N_{orb}\times N_{orb}$) is
$8N_{orb}^3$.
Thus, the overall flop count
for OMEN SSE is $64N_aN_bN_{3D}N_{k_z}N_{q_z}N_EN_{\omega}N_{orb}^3$.
The multiplication reduction (algebraic regrouping) powered by the data-centric view (\S~\ref{sec:dataflow}) decreases it by
$\frac{2N_{q_z}N_\omega}{N_{q_z}N_\omega+1}$, essentially half of the
flops for practical sizes.

\begin{table}[t]
\caption{Single Iteration Computational Load$^*$ (in Pflop)}
\small
\vspace{-1em}
\label{tab:flop}
\begin{tabular}{ l rrrrr } 
 \toprule
 & \multicolumn{5}{c}{$\boldsymbol{N_{k_z}}$}\\\cmidrule{2-6}
 \textbf{Kernel} & 3 & 5 & 7 & 9 & 11 \\
 \midrule
Boundary Conditions & 8.45 & 14.12 & 19.77 & 25.42 & 31.06\\
RGF & 52.95 & 88.25 & 123.55 & 158.85 & 194.15\\
SSE (OMEN) & 24.41 & 67.80 & 132.89 & 219.67 & 328.15\\
SSE (DaCe) & 12.38 & 34.19 & 66.85 & 110.36 & 164.71\\
\bottomrule
\end{tabular}\\\vspace{0.3em}
{\footnotesize $^*$: ``Small'' structure.}\vspace{-1em}
\end{table}

\subsubsection{Communication Model}\label{sec:commmodel}
Computing ${\mathbf \Sigma}^\gtrless\left(k_z, E\right)$ in SSE req-uires each electron pair to execute the following $N_{q_z}N_\omega$ times:
\begin{itemize}
\item Receive the phonon Green's Functions ${\mathbf D}^\gtrless(q_z, \omega)$;
\item Receive the electron Green's Functions ${\mathbf G}^\gtrless(k_z-q_z, E\pm\omega)$.
Symmetrically, send its own electron data ${\mathbf G}^\gtrless(k_z, E)$ to the
$\left(k_z+q_z, E\mp\omega\right)$ points;
\item Accumulate a partial sum of the interactions of the atoms with its neighbors,
as described by Eq.~(\ref{eq:3}).
\end{itemize}
We note that the computation of ${\mathbf \Pi}^\gtrless$ follows a similar pattern.
Putting it all together, the communication scheme for SSE in OMEN is split into $N_{q_z}N_{\omega}$ rounds.
In each round:
\begin{itemize}
\item The phonon Green's Functions ${\mathbf D}^\gtrless(\omega, q_z)$ are broadcast to all electron processes;
\item Each electron process iterates over its assigned electron Green's Functions ${\mathbf G}^\gtrless(E, k_z)$
and receives the corresponding ${\mathbf G}^\gtrless(E\pm\hbar\omega, k_z-q_z)$.
In a symmetrical manner, it sends its assigned Green's Functions to all $\left(k_z+q_z, E\mp\omega\right)$ points;
\item The partial phonon self-energies $\Pi_p^\gtrless(\omega, q_z)$ produced by each electron process are reduced
to ${\mathbf \Pi}^\gtrless(\omega, q_z)$ and sent to the corresponding phonon process. 
\end{itemize}
Based on the above, we make the following observations:
\begin{itemize}
\item All ${\mathbf D}^\gtrless$ are broadcast to all electron processes;
\item All ${\mathbf G}^\gtrless$ are replicated through point-to-point communication $2N_{q_z}N_{\omega}$ times.
\end{itemize}
To put this into perspective, consider a ``Large'' structure simulation with $N_E=$ 1,000. The communication aspect of the SSE
phase involves \textbf{each electron process receiving and sending 276 GiB for ${\mathbf D}^\gtrless$ (${\mathbf \Pi}^\gtrless$),
as well as 2.58 PiB for ${\mathbf G}^\gtrless$, independent of the number of processes}.

In the transformed scheme (Fig.~\ref{fig:ssetiling}, right), each process is assigned a subset of $\frac{N_a}{T_a}$ atoms
and $\frac{N_E}{T_E}$ electron energies,
where $P=T_aT_E$ is the number of processes.
The computations described in Eqs.~(\ref{eq:3})--(\ref{eq:4}) require all neighbors per atom, which may or may not be in the same subset. Therefore, the actual
number of atoms that each process receives is $\frac{N_a}{T_a} + c$, where $c$
is the number of neighboring atoms that are not part of the subset. We over-approximate $c$ by $N_b$.
In a similar manner, each process is assigned $\frac{N_E}{T_E} + 2N_\omega$ energies.

We implement the distribution change with all-to-all collective operations
(\texttt{MPI\_Alltoallv} in the MPI standard).
We use four collectives on ${\mathbf G}^\gtrless$, ${\mathbf D}^\gtrless$, ${\mathbf \Sigma}^\gtrless$, and ${\mathbf \Pi}^\gtrless$,
where each process contributes:
\begin{itemize}
\item $64N_{k_z}\left(\frac{N_E}{T_E}+2N_\omega\right)\left(\frac{N_a}{T_a}+N_b\right)N_{orb}^2$ bytes for  ${\mathbf G}^\gtrless$ and ${\mathbf \Sigma}^\gtrless$;
\item $64N_{q_z}N_{\omega}\left(\frac{N_a}{T_a}+N_b\right)(N_b+1)N_{3D}^2$ bytes for ${\mathbf D}^\gtrless$ and ${\mathbf \Pi}^\gtrless$.
\end{itemize}

We quantify the communication volumes for a typical problem size in Tables \ref{tab:weak-bytes} and \ref{tab:strong-bytes}, the former with varying $N_{k_z}$ and the latter with fixed parameters. Both tables highlight a clear advantage in favor of the communication-avoiding variant, communicating two orders less than the state of the art.
For the large-scale simulation described above, the new communication approach
(with $T_a = P, T_E = 1$) distributes
the 276 GiB for ${\mathbf D}^\gtrless$ and ${\mathbf \Pi}^\gtrless$ over \textit{all} processes,
and only adds a minor overhead of 28.26 MiB per process. It also lowers the fixed cost of 2.58 PiB for
${\mathbf G}^\gtrless$ and ${\mathbf \Sigma}^\gtrless$ to only 1.8 TiB distributed
to all processes and 6.13 GiB per process. We note that the total cost for ${\mathbf G}^\gtrless$
becomes equal for the two communication schemes when the number of processes is
greater than 440,000.

\begin{table}
    \setlength{\tabcolsep}{3pt}
	\begin{center}
    \caption{SSE Communication Volume$^*$ Weak Scaling (TiB)}\vspace{-1em}
		\label{tab:weak-bytes}
		\small		
		\begin{tabular}{l lllll} 
			\toprule
			 & \multicolumn{5}{c}{\bf $\boldsymbol{N_{k_z}}$ (Processes)}\\\cline{2-6}
			\textbf{Variant} &3 (768)&5 (1280)&7 (1792)&9 (2304)&11 (2816)\\
			\midrule
			OMEN & 32.11 & 89.18 & 174.80 & 288.95 & 431.65\\
			DaCe & 0.54 [59$\times$] & 1.22 [73$\times$] & 2.17 [81$\times$] & 3.38 [85$\times$] & 4.86 [89$\times$]\\
			\bottomrule
		\end{tabular}\\\vspace{0.3em}
		{\footnotesize $^*$: ``Small'' structure, reduction ratio in brackets.}
	\end{center}
	\vspace{-1em}
\end{table}
\begin{table}[t]
    \setlength{\tabcolsep}{3pt}
	\begin{center}
    \caption{SSE Communication Volume$^*$ Strong Scaling (TiB)}\vspace{-1em}
		\label{tab:strong-bytes}
		\small		
		\begin{tabular}{l lllll} 
			\toprule
			& \multicolumn{5}{c}{\bf Processes}\\\cline{2-6}
			\textbf{Variant} & 224 & 448 & 896 & 1792 & 2688 \\
			\midrule
			OMEN & 108.24 & 117.75 & 136.76 & 174.80 & 212.84\\
			DaCe & 0.95 [114$\times$] & 1.13 [104$\times$] & 1.48  [92$\times$] & 2.17 [80$\times$] & 2.87 [74$\times$]\\
			\bottomrule
		\end{tabular}\\\vspace{0.3em}
		{\footnotesize $^*$: ``Small'' structure, $N_{k_z}=7$, reduction ratio in brackets.}
	\end{center}
	\vspace{-1em}
\end{table}

\subsection{Selected HPC Platforms}

The two systems we run DaCe OMEN experiments on are CSCS Piz Daint~\cite{daint} (6th place in June 2019's Top500 supercomputer list~\cite{hpl}) and OLCF Summit~\cite{summit} (1st place). 
Piz Daint is composed of 5,704 Cray XC50 compute nodes, each equipped with a
12-core HT-enabled (2-way SMT) Intel Xeon E5-2690 v3 CPU with 64 GiB RAM (peaking at 499.2 double-precision Gflop/s), 
and one NVIDIA Tesla P100 GPU (4.7 Tflop/s). The nodes communicate using Cray's Aries interconnect.
Summit comprises 4,608 nodes, each containing two IBM POWER9 CPUs (21 usable physical cores with 4-way SMT, 515.76 Gflop/s) with 512 GiB RAM and six NVIDIA Tesla V100 GPUs (42 double-precision Tflop/s in total, 720 half-precision Tflop/s using Tensor Cores). The nodes are connected using Mellanox EDR 100G InfiniBand organized in a Fat Tree topology.
For Piz Daint, we run our experiments with two processes per node (sharing the GPU),
apart from a full-scale run on 5,400 nodes, where the simulation parameters do not
produce enough workload for more than one process per node.
In Summit we run with six processes per node, each consuming 7 physical cores.
We conduct experiments at least 5 times, reporting the median
and 95\% Confidence Interval.

Despite the fact that both systems feature GPUs as their main workhorse,
the rest of the architecture is quite different.
While Piz Daint has a reasonable
balance between CPU and GPU performance (GPU/CPU ratio of 9.4$\times$), Summit's POWER9 CPUs are significantly (81.43$\times$) weaker than the six V100 GPUs present on each node. 
Tuning our data-centric simulator, which utilizes both the CPU and GPUs, we 
take this into account by assigning each architecture different tile sizes
and processes per node, so as to balance the load without running out of memory.

\section{Performance Results}\label{sec:results}

We proceed to evaluate the performance of the data-centric OMEN algorithm. 
Starting with per-component benchmarks, we demonstrate the necessity of specialized 
implementations, and that critical portions of the algorithm
are sufficiently tuned to the underlying systems. 
We then measure performance aspects of OMEN and the DaCe variant on the ``Small'' problem, consisting of 4,864 atoms,
on up to 5,400 Piz Daint and 1,452 Summit GPUs.
Lastly, we measure the heat dissipation of the ``Large'', 10,240 atom nanodevice on up to 27,360 Summit GPUs.

\subsection{Component Benchmarks}

Below we present individual portions of the quantum transport simulation pipeline, including data I/O, overlapping, computational aspects of GF, SSE, and total single node performance.

\subsubsection{Data Ingestion}

The input of the simulator is material and structural information of the nano-device in question (produced by CP2K). The size of this data typically ranges between the order of GiBs to 10s of GiBs, scattered across multiple files.
Once the data is loaded and pre-processed for each rank, the OMEN algorithm does not require additional I/O, and can operate with the information dispersed across the processes. 
Despite being a constant overhead, without proper data staging, running at large scale quickly becomes infeasible due to contention on the parallel filesystem. For example, running on Piz Daint at near-full scale (5,300 nodes) takes over 30 minutes of loading, and with 2,589 nodes takes 1,112 seconds.

To reduce this time, we stage the nanostructure information and broadcast relevant information in chunks to the participating ranks. As expected, this reduces start-up time to under a minute in most cases, and 31.1 seconds for 4,560 nodes.
As the solver normally takes 20--100 iterations to converge, we also report time-per-iteration factoring for I/O overhead w.r.t. 50 iterations.

\subsubsection{Recomputation Avoidance} \label{sec:caching}
For each iteration of the Green's Function phase and for each energy-momentum point,
three operations are performed: (a) specialization of the
input data, (b) computation of the boundary conditions, and (c) execution of the
RGF kernel. From a data-centric perspective, the first two operations induce data dependencies on specific energy-momentum points, but not on the iteration. Therefore, it is possible to perform them once
during an initialization phase and reuse the produced output in all iterations.
However, the memory footprint to avoid recomputation is immense: specialization data for the ``Large'' nanodevice consumes 3GB per-point, while the boundary conditions need another 1GB. We leverage the compute-memory tradeoff by enabling three modes of executing the GF phase:
(a) ``No Cache'': data is recomputed in every iteration; (b) ``Cache Boundary Conditions'', where data is only re-specialized per-iteration; and (c) ``Cache BC+Spec.'': both specialized data and boundary conditions are cached in memory.

\subsubsection{Automatic Pipelining}
Since parallelism is expressed natively in SDFGs, the DaCe framework schedules independent nodes concurrently, using threads for CPU and CUDA streams for GPU.
DaCe automatically generates 32 and 60 streams for electron and phonon GF computations, respectively.
We vary the maximum number of concurrent streams on a single electron point on Summit and report the results in Table \ref{tab:pipe}. The results imply that more than 16 streams are necessary to yield the 7.5\% speedup in the table, automatically achieved as a byproduct of data-centric parallel programming.

\begin{table}[h]
	\vspace{-0.5em}
	\caption{CUDA Streams in Green's Functions (Summit)}\vspace{-1em}
	\label{tab:pipe}
	\small
	\begin{tabular}{llllll}
		\toprule
		\textbf{Streams}   & 1 & 2 & 4 & 16 & Auto (32)\\\midrule 
		\textbf{Time [s]}  & 10.07  & 9.94  & 9.86  & 9.61 & \textbf{9.32} (2.63 Tflop/s) \\
		\bottomrule
	\end{tabular}\vspace{-1em}
\end{table}

\subsubsection{Green's Functions and Sparsity}

Since the RGF algorithm uses a combination of sparse and dense matrices, 
several alternatives exist to perform the required multiplications.
In particular, a common operation in
RGF is \texttt{F[n] @ gR[n + 1] @ E[n + 1]} --- multiplying two blocks ($E$, $F$) of the
block tri-diagonal Hamiltonian matrix ($H$) with a retarded Green's Functions block ($gR$).
The off-diagonal blocks are typically sparse CSR matrices, but can also be stored in CSC format if needed.
To perform this operation, one can either use CSR-to-dense
conversion followed by dense multiplication (\textit{GEMM}); or use sparse-dense multiplications (\textit{CSRMM}).
These options can be interchanged via data-centric transformations.

\begin{table}[h]
	\vspace{-0.75em}
	\setlength{\tabcolsep}{4pt}
	\caption{Matrix Multiplication Performance}\vspace{-1em}
	\label{tab:gemm}
	\small
	\begin{tabular}{lrrrrr}
		\toprule
		& & \multicolumn{4}{c}{\textbf{Operation}}\\
		\cmidrule(l{2pt}r{2pt}){3-6}
		\textbf{GPU} & \textbf{Method} & \textbf{NN} & \textbf{NT} & \textbf{TN} & \textbf{TT}\\
		\midrule
		\multirow{2}{*}{P100} & GEMM & 100.337 ms & 99.306 ms & 101.959 ms & 100.857 ms \\
							  & CSRMM2 & 15.937 ms & \textbf{11.437} ms & 85.573 ms & --- \\
							  & GEMMI & 30.861 ms & --- & --- & --- \\
		\multirow{2}{*}{V100} & GEMM & 58.382 ms & 58.144 ms & 58.666 ms & 58.315 ms \\
							  & CSRMM2 & 8.202 ms & \textbf{6.14} ms & 52.722 ms & --- \\
							  & GEMMI & 15.198 ms & --- & --- & --- \\
		
		\bottomrule
	\end{tabular}\vspace{-1.25em}
\end{table}

In Table~\ref{tab:gemm} we study the performance of the different methods
in cuBLAS and cuSPARSE, on the P100 and V100 GPUs.
The \textit{CSRMM2} method multiplies a CSR matrix (on the left side) with a dense matrix and supports NN, NT, and TN operations.
The \textit{GEMMI} method multiplies a dense matrix with a CSC matrix (on the right side) and only supports NN.
In Table~\ref{fig:mb:gf} we study the performance of three different approaches to
compute the above product.
In the first approach, we use dense multiplication twice.
In the second approach, we assume that both $F$ and $E$ are CSR matrices. We first compute the product
of $E$ and $gR$ with CSRMM2 in TN operation. Subsequently, we multiply the intermediate
result with $F$ using GEMMI.
In the third approach, $F$ is in CSR format while $E$ is given in CSC. We compute the product
of $F$ and $gR$ with CSRMM2 in NT operation. Using a second identical operation
we multiply $E$ with the intermediate result.
We observe that the best performance is attained with the third approach, with
5.10--9.74$\times$ speedup over other GPU implementations.

\begin{table}[h]
	\vspace{-1em}
	\setlength{\tabcolsep}{3pt}
	\caption{3-Matrix Multiplication Performance}\vspace{-1em}
	\label{fig:mb:gf}
	\small
	\begin{tabular}{lrrr}
		\toprule
		& \multicolumn{3}{c}{\textbf{Approach}}\\
		\cmidrule(l{2pt}r{2pt}){2-4}
		\textbf{GPU} & \textbf{GEMM/GEMM} & \textbf{CSRMM2/GEMMI} & \textbf{CSRMM2/CSRMM2}\\
		\midrule
		P100 & 200.879 ms & 116.380 ms & \textbf{22.798} ms \\
		V100 & 116.881 ms & 67.924 ms & \textbf{11.994} ms \\		
		\bottomrule
	\end{tabular}\vspace{-1.25em}
\end{table}

\begin{figure}[t]
	\begin{subfigure}{\linewidth}
		\centering
		\includegraphics[width=.85\linewidth]{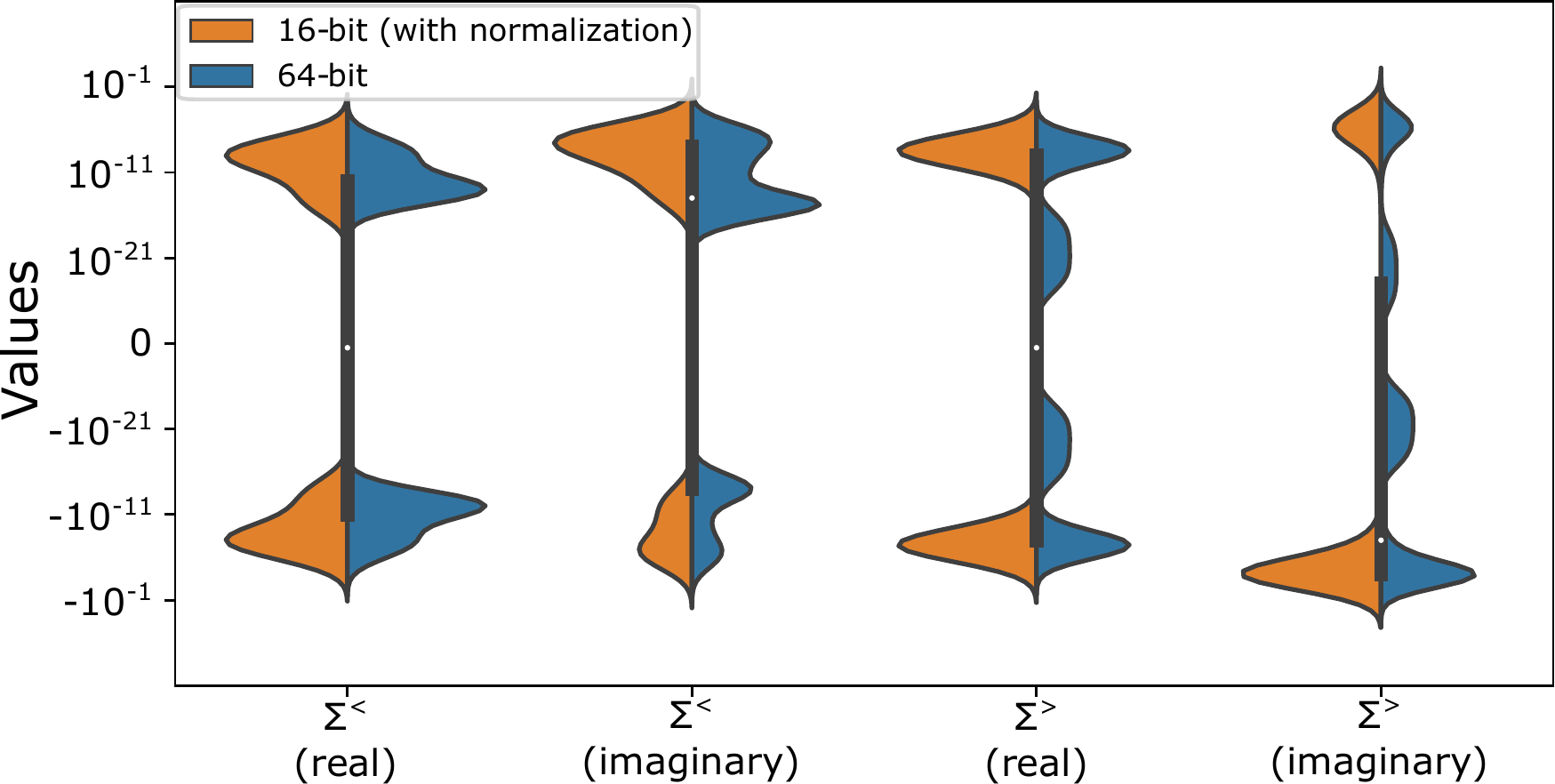}
		\caption{Output Distribution (Non-Zero Values)}
		\label{fig:histogram}
	\end{subfigure}
	\begin{subfigure}{\linewidth}
		\centering
		\includegraphics[width=.87\linewidth]{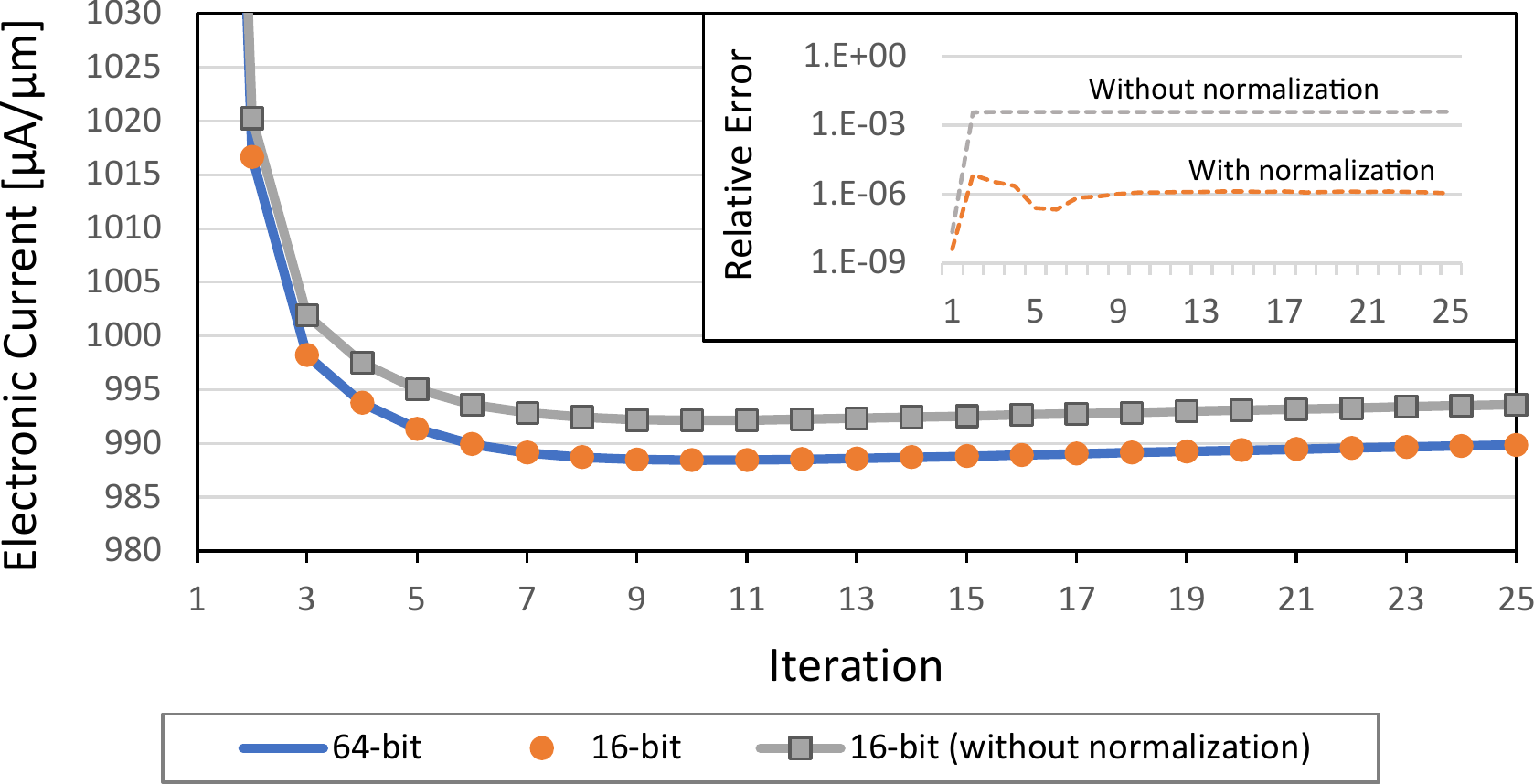}
		\caption{Convergence}
		\label{fig:convergence}
	\end{subfigure}
	\vspace{-1em}
	\caption{Comparison of double- and half-precision SSE.}
	\vspace{-1.25em}
\end{figure}
\begin{figure*}
	\begin{subfigure}[b]{.42\linewidth}
		\centering
		\includegraphics[width=\linewidth]{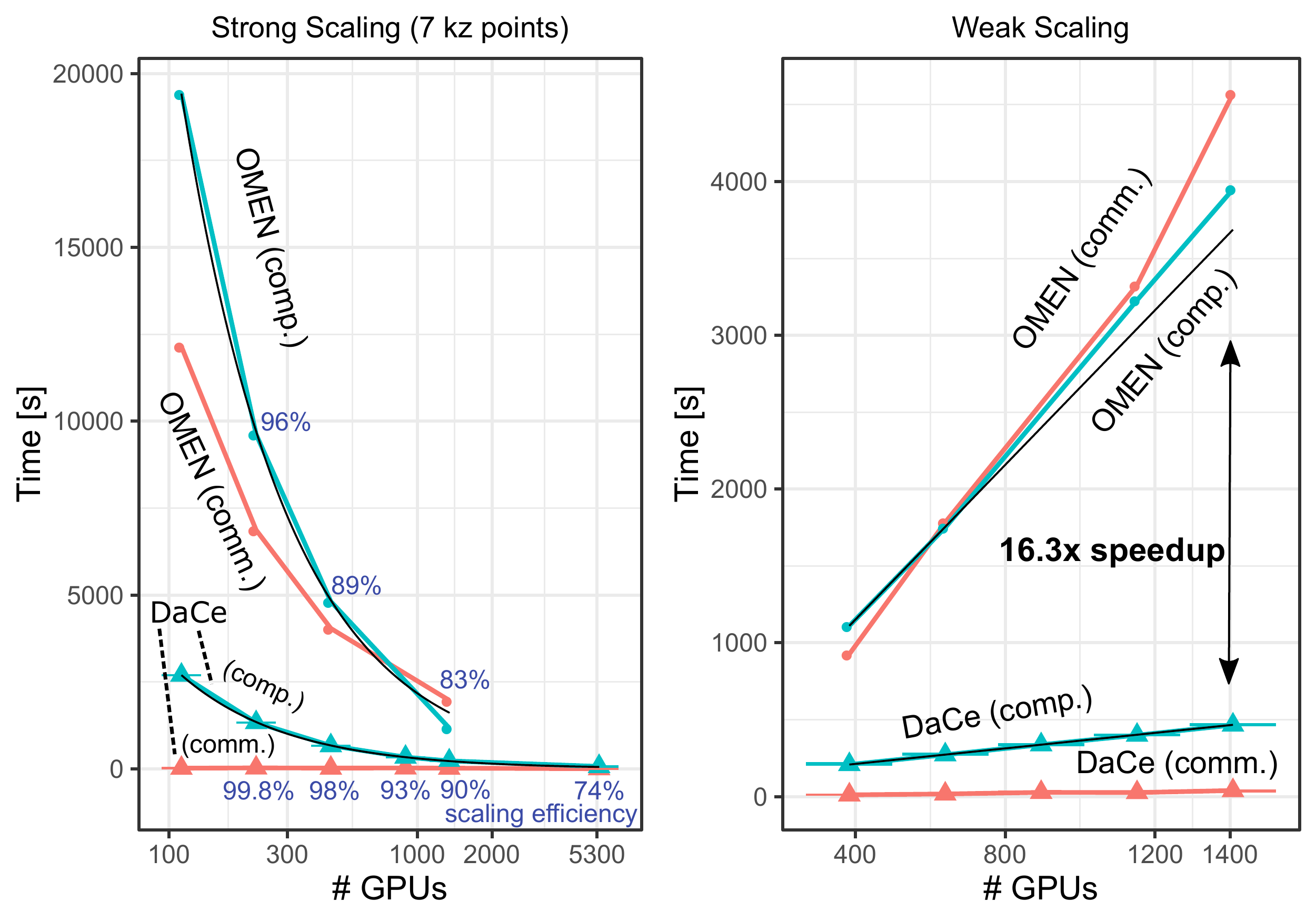}
		\vspace{-2em}
		\caption{Piz Daint}		
	\end{subfigure}\qquad\qquad
	\begin{subfigure}[b]{.42\linewidth}
		\centering
		\includegraphics[width=\linewidth]{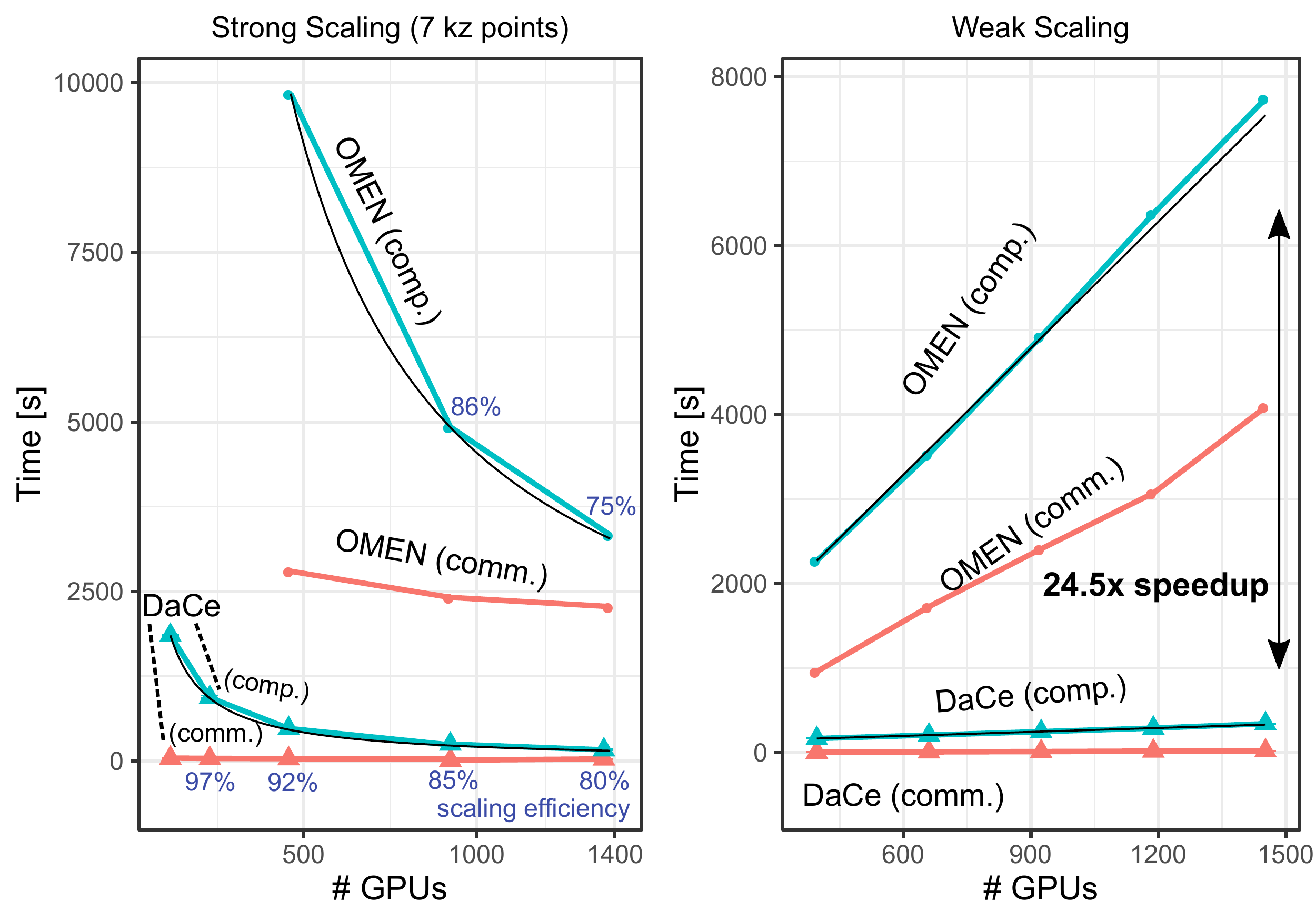}
		\vspace{-2em}
		\caption{Summit}		
	\end{subfigure}
	\vspace{-1em}
	\caption{DaCe OMEN simulation scalability ($N_a=4{,}864$, black lines:
		ideal scaling).}
	\vspace{-1em}
	\label{fig:scalability}
\end{figure*}

\subsubsection{Custom Strided Multiplication}\label{sec:sbsmm}
As part of SSE dataflow transformations (\S~\ref{sec:dataflow}), we 
reformulate a multitude of small-scale matrix multiplications into one ``strided 
and batched'' GEMM operation (Fig.~\ref{fig:transformation}, step \three). 
We thus initially opt to use the highly-tuned NVIDIA cuBLAS library for the computation, which yields 85.7\% of peak double-precision flop/s on average. 
However, upon deeper inspection using the performance model (\S~\ref{sec:compmodel}), we observe a discrepancy, where the useful operations/second with respect to the actual sizes is only around 6\% of the GPU peak performance. 
As our individual matrices are small (typically 12$\times$12), this effect may stem from \textit{excessive padding} in cuBLAS, which is tuned specifically for common problem sizes.

\begin{table}[h]
	\vspace{-0.75em}
	\setlength{\tabcolsep}{4pt}
	\caption{Strided Matrix Multiplication Performance}\vspace{-1em}
	\label{tab:sbmm}
	\small
	\setlength{\tabcolsep}{0.4em}
	\begin{tabular}{lrrrrrr}
		\toprule
		& \multicolumn{3}{c}{\textbf{cuBLAS}} & \multicolumn{3}{c}{\textbf{DaCe (SBSMM)}}\\
		\cmidrule(l{2pt}r{2pt}){2-4}\cmidrule(l{2pt}r{2pt}){5-7}
		\textbf{GPU} & Gflop & Time & \% Peak (Useful) & Gflop & Time & \% Peak \\
		\midrule
		P100 & 27.42 & 6.73 ms & 86.6\% (6.1\%) & 1.92 & 4.03 ms & 10.1\% \\
		V100 & 27.42 & 4.62 ms & 84.8\% (5.9\%) & 1.92 & 0.70 ms & 39.1\% \\
		V100-TC & --- & --- & --- & 3.42 & 0.13 ms & --- \\
		
		\bottomrule
	\end{tabular}\vspace{-1em}
\end{table}

We use DaCe to create two \textit{specialized} strided-batched small-scale
matrix multiplication tasklets (SBSMM in Fig.~\ref{fig:transformation}),
and report their performance in Table \ref{tab:sbmm}. The double-precision tasklet maximizes parallelism based on the problem size and does not pad data; whereas the half-precision tasklet utilizes Tensor Cores (\S \ref{sec:mixedprec}) for complex matrix multiplications. \emph{As shown in the
table, SBSMM is 5.76$\times$ (64-bit) to 31$\times$ (16-bit) faster than cuBLAS, which 
does not support half-complex multiplication, demonstrating that performance engineers 
can use the data-centric view to partition specific problems in ways not considered by 
vendor HPC libraries.}

\subsubsection{Mixed-Precision Convergence}
Figures \ref{fig:histogram} and \ref{fig:convergence} depict the output value distribution and convergence of the electronic current for SSE and  the reduced-precision scheme (SSE-16, \S~\ref{sec:mixedprec}) for the ``Large'' structure. When normalization is applied, SSE-16 produces similar outputs per step and converges at the same rate as SSE, to a value that relatively differs by $1.2\cdot 10^{-6}$. In comparison, without scaling $D^\gtrless,G^\gtrless$, the relative error increases to 0.003. The core multiplication kernels, SBSMM, yield 14.78 Tflop/s per-GPU, \textbf{404.41 Pflop/s} in total over 4,560 Summit nodes.

\subsubsection{Single-Node Performance}\label{sec:nodeperf}

We evaluate the performance of OMEN, the DaCe variant, and the Python 
implementation (using the numpy module implemented over MKL), on the ``Small'' Silicon nanostructure with $N_{k_z} = 3$.
Table~\ref{tab:single-node} shows the runtime of the GF and SSE SDFG
states,
for $\frac{1}{384}$ of the total computational load, executed by a single node on Piz Daint. Although Python uses optimized routines, it exhibits very slow performance on its own. This is a direct result of utilizing an interpreter for mathematical expressions, where
arrays are allocated at runtime and each operation incurs high overheads. This can especially be seen in SSE, which consists of many small multiplication operations. 
The table also indicates that the data-centric transformations made on the Python code
using DaCe outperforms the manually-tuned C++ OMEN on both phases, where the performance-oriented reconstruction of SSE generates a speedup of 9.97$\times$.

\begin{table}[h]
	\begin{center}
    \caption{Piz Daint Single-Node Performance\vspace{-1em}}
		\small
		\label{tab:single-node}
		\begin{tabular}{l rrrrrr} 
			\toprule
            & \multicolumn{6}{c}{\bf Phase}\\
            \cline{2-7}\addlinespace
            \multirow{2}{*}{\bf Variant} & \multicolumn{3}{c}{GF} & \multicolumn{3}{c}{SSE} \\
            \cmidrule(l{2pt}r{2pt}){2-4}\cmidrule(l{2pt}r{2pt}){5-7}
			
            & Tflop & Time [s] & \% Peak & Tflop & Time [s] & \% Peak \\
            \midrule
            OMEN & 174.0 & 144.14 & 23.2\% & 63.6 & 965.45 & 1.3\% \\
            Python & 174.0 & 1,342.77 & 2.5\% & 63.6 & 30,560.13 & 0.2\% \\
            DaCe & 174.0 & \textbf{111.25} & \textbf{30.1\%} & 31.8 & \textbf{29.93} & \textbf{20.4\%} \\
			\bottomrule
		\end{tabular}
	\end{center}
	\vspace{-1.5em}
\end{table}

\subsubsection{Communication} \label{sec:comm_eff}

We study DaCe OMEN's communication efficiency on the Summit supercomputer.  We
utilize \texttt{MPI\_Alltoallv} to exchange $D^\gtrless$,
$\Pi^\gtrless$, $G^\gtrless$, and $\Sigma^\gtrless$.  In this collective call,
each rank sends a different amount of data to all other ranks, performed in several rounds~\cite{prisacari2013bandwidth}.  For OMEN, the
communication pattern is sparse for some of the calls. We derive lower bounds
for the completion time of each call by counting the amount of data each
node must send (aggregating over all ranks on the same node), and dividing that
by the injection bandwidth of a Summit node of 23 GB/s \cite{summit}.

For the large-scale run, our model predicts 1.85s of runtime to communicate each of $D^\gtrless$/$\Pi^\gtrless$
and 0.21s to communicate each of $G^\gtrless$/$\Sigma^\gtrless$ at 100\% of
injection bandwidth utilization. Our measurements (\S~\ref{sec:extreme}) show that
we achieve 84.57\% and 42.32\% of that (2.18s and 0.55s actual runtime for $D^\gtrless$/$\Pi^\gtrless$
and $G^\gtrless$/$\Sigma^\gtrless$, respectively).

\subsection{Scalability}

The communication-avoiding variant of OMEN (\textit{DaCe OMEN}) scales
well to the full size of both supercomputers. In Fig.~\ref{fig:scalability}, we
measure the runtime and scalability of a single GF-SSE iteration of OMEN and
the DaCe variant on Piz Daint and Summit. 
For strong scaling, we use the ``Small'' structure and fix $N_{k_z}=7$ (so that OMEN can treat 
it), running with 112--5,400 nodes on Piz Daint and 19--228 nodes (114--1,368 GPUs) on Summit.
For weak scaling, we annotate ideal scaling (in black) with proportional
increases in the number of $k_z$ points and nodes, since
the GF and SSE phases scale differently relative to the simulation parameters, by $N_{k_z}$ and $N_{k_z}N_{q_z}=N_{k_z}^2$, respectively.
We measure the same structure with varying $k_z$ points: $N_{k_z}\in\left\{3,5,7,9,11\right\}$, using 384--1,408 nodes on Piz Daint and 
66--242 nodes (396--1,452 GPUs) on Summit.

Compared with the original OMEN, the DaCe variant is efficient, both from the
computation and communication aspects. On Piz Daint, the total runtime of the reduced-communication
variant \textbf{outperforms OMEN, the current state of the art, up to a factor of 16.3$\times$},
while the communication time \textbf{improves by up to 417.2$\times$}. On Summit,
the total runtime \textbf{improves by up to factor of 24.5$\times$}, while communication
is \textbf{sped up by up to 79.7$\times$}. Moreover, the higher the simulation accuracy ($N_{k_z}$), the greater the speedup is. 

The speedup of the computational runtime on Summit is higher than on Piz Daint.
This is the result of OMEN depending on multiple external libraries, some of which are not
necessarily optimized for every architecture (e.g., IBM POWER9). On the other hand, 
SDFGs are compiled on the target architecture and depend only on a few optimized 
libraries provided by the architecture vendor (e.g., MKL, cuBLAS, ESSL), whose implementations can always be replaced by SDFGs for further tuning and transformations. 

As for scaling efficiency, on Summit DaCe OMEN achieves a speedup of 9.68$\times$ on 12 times the nodes in the strong scaling experiment (11.23$\times$ for computation alone). Piz Daint yields similar results with 10.69$\times$ speedup. 
The algorithm weakly scales with $N_{k_z}$ on both platforms, again an
order of 
magnitude faster than the state of the art on structures of the same size. We  thus conclude that the data-centric variant of OMEN is strictly desirable over the original. 

\vspace{-0.5em}
\subsection{Extreme-Scale Runs}\label{sec:extreme}

We run DaCe OMEN on a setup not possible on the original OMEN,
due to infeasible memory requirements. We simulate the ``Large'' 
10,240 atom nanostructure --- \textit{a size 
never-before-simulated with GF+SSE at the ab initio level} --- using the 
DaCe variant of OMEN. For this purpose, we use up to 98.96\% of the Summit 
supercomputer, 27,360 GPUs, and run our proposed Python code with 21 $k_z$ points, 
which are necessary to produce accurate results (see Table 
\ref{tab:common-values}).

\begin{figure}
	\centering
	\includegraphics[width=.9\linewidth]{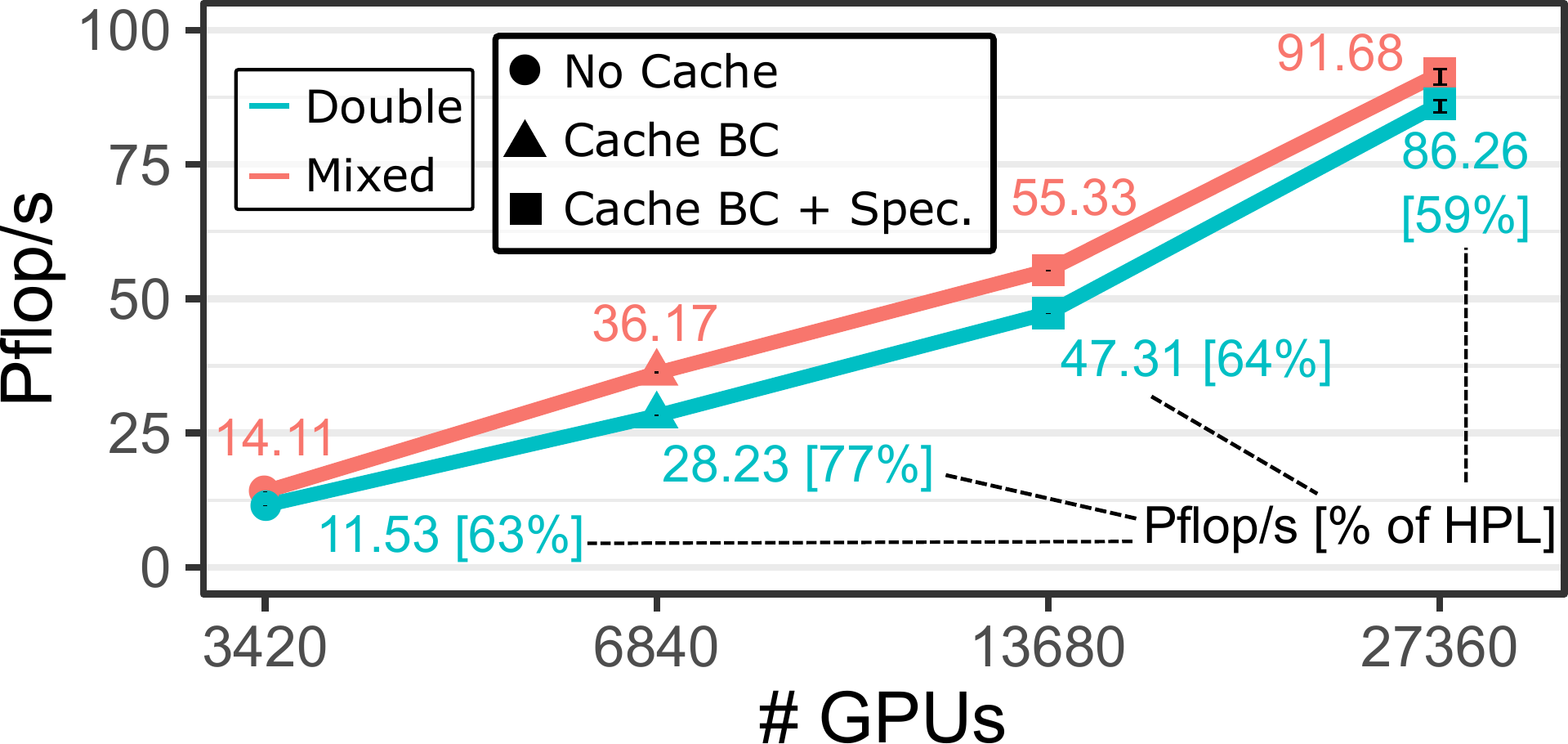}
	\vspace{-1em}
	\caption{Strong scaling on Summit, ``Large'' structure.}
	\vspace{-1em}
	\label{fig:extreme}
\end{figure}

Figure \ref{fig:extreme} plots the results of the strong-scaling experiment, 
using 3,420 GPUs for the baseline. The simulation costs 8.17--9.41 Exaflop per 
iteration, depending on the caching strategy (specialization and/or boundary conditions, \S~\ref{sec:caching}).
Sustained performance of \textbf{85.45 Pflop/s (42.55\% of supercomputer peak)} is
achieved in double precision, and \textbf{90.89.68 Pflop/s} in mixed precision,
including communication, I/O, and caching as described above. A full breakdown is listed in Table 
\ref{tab:breakdown}. The table compares performance with machine peak and effective maximum performance (HPL, 148.6 Pflop/s for Summit \cite{hpl}).
Additionally, we compare the per-atom performance of the DaCe variant with the
original OMEN on 6,840 Summit GPUs. Both implementations execute a simulation
with 21 $k_z$ points and 1,220 electron energies, but different number of atoms.
As shown in Table~\ref{tab:two-orders}, DaCe OMEN is up to \textbf{two orders of magnitude
faster per-atom}.
These results prove that the electro-thermal properties 
of nano-devices of this magnitude can be computed in under 2 minutes per 
iteration, as desired for practical applications.

\begin{table}[h]
	\vspace{-0.5em}
  \caption{Full-Scale 10,240 Atom Run Breakdown}\vspace{-1em}
	\label{tab:breakdown}
	\small
	\setlength{\tabcolsep}{0.5em}
	\begin{tabular}{lrrrrr}
		\toprule
		Phase & Time & Eflop & Pflop/s & \% Max & \% Peak\\\midrule
		Data Ingestion & 31.10 & --- & --- & --- & ---\\
		Boundary Conditions & 30.51 & 1.23 & 40.40 & 27.19\% & 20.12\%\\
		\midrule
		GF & 41.36 & 6.00 & 145.01 & 97.59\% & 72.22\%\\
		SSE (double-precision) & 41.91 & 2.18 & 51.94 & 34.95\% & 25.87\%\\
		SSE (mixed-precision) & 36.16 & 2.18 & 60.21 & --- & ---\\
		Communication & 11.50 & --- & --- & --- & 76.72\%\\
		\midrule
		\textbf{Total} & 94.77 & 8.17 & 86.26 & 58.05\% & 42.96\%\\
		\textbf{Total} (incl. I/O and Init.) & 96.00 & 8.20 & 85.45 & 57.50\% & 42.55\%\\
		\midrule
		\textbf{Total (mixed-precision)} & 89.02 & 8.17 & 91.68 & --- & ---\\
		\textbf{Total} (incl. I/O and Init.) & 90.25 & 8.20 & 90.89 & --- & ---\\
		\bottomrule
	\end{tabular}\vspace{-1.5em}
\end{table}
\begin{table}[h]
    \caption{Per-Atom Performance}\vspace{-1.5em}
      \label{tab:two-orders}
      \small
      \begin{tabular}{lrrrr}
          \toprule
          Variant & $N_a$ & Time [s] & Time/Atom [s] & Speedup\\
          \midrule
          OMEN & 1,064 & 4,695.70 & 4.413 & 1.0x\\
          DaCe & 10,240 & 333.36 & 0.033 & 140.9x\\
          \bottomrule
      \end{tabular}
      $P= 6{,}840, N_b=34, N_{orb} = 12, N_E = 1{,}220, N_{\omega}=70$.\vspace{-0.5em}
  \end{table}

\begin{figure}
	\centering
	\includegraphics[width=.72\linewidth]{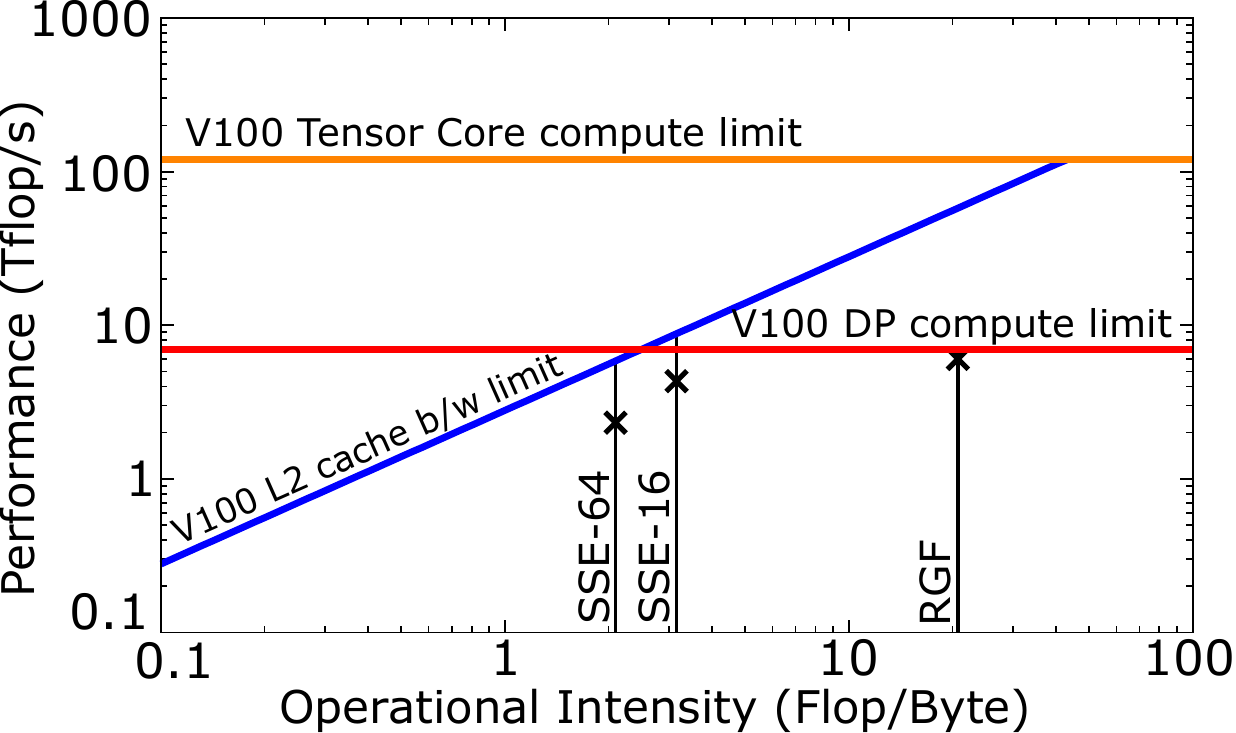}
	\vspace{-1em}
	\caption{Roofline model of the computational kernels.}
	\vspace{-1em}
	\label{fig:roofline}
\end{figure}

To gain insight on the factors that limit the performance of DaCe OMEN, we analyze
the bottlenecks of each phase in Table \ref{tab:breakdown}. We use the Roofline
model \cite{roofline} to depict the limits of the main phases in Fig.~\ref{fig:roofline}.
The GF phase, as seen in Fig.~\ref{fig:roofline},
is compute-bound, achieving 97.59\% of HPL performance. In contrast, the SSE phase
combines a multitude of small matrix multiplications, which were shown to be
memory-bound~\cite{Masliah}. This agrees with our empirical results, in which
the memory per batched multiplication is small enough to fit in the L2 cache,
and operational intensity is too low to be compute-bound. Mixed-precision SSE
(Fig.~\ref{fig:roofline}, SSE-16) improves I/O by reducing element size, but is
still limited by memory bandwidth from a data-centric perspective. Communication
(\S~\ref{sec:comm_eff} for a full model) is a sparse alltoall collective,
which achieves in total 76.72\% bandwidth utilization and cannot be overlapped with
computation algorithmically.

\begin{figure*}
	\centering
	\includegraphics[width=.75\linewidth]{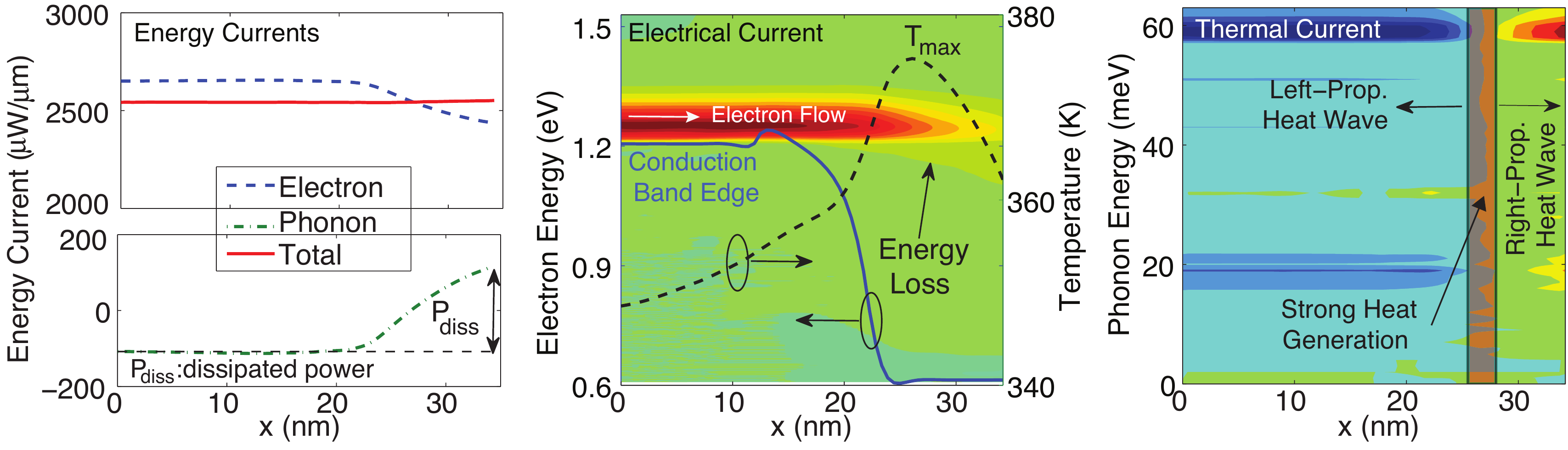}
	\vspace{-1em}
	\caption{Electro-thermal simulation of a FinFET with $W$=2.1nm
          and $L$=35nm (``Small'' structure from Section
          \ref{omen_struct}, same as in Fig.~\ref{fig:res}). The energy (left), electrical (middle), and thermal
          (right) currents are shown at $V_{gs}$=0.5 V and $V_{ds}$=0.6 V.}
	\vspace{-1em}      
	\label{fig:res_2}
\end{figure*}

\section{Implications}

\subsection{Quantum Transport Simulations}\label{sec:qt_sim}
With DaCe OMEN, the electro-thermal properties of FinFET-like
structures can now be simulated within ultra-short times. An example
is shown in Fig.~\ref{fig:res_2}. The left sub-plot represents the
electron (dashed blue line) and phonon (dashed-dotted green line)
energy currents that flow through the considered device. As their sum
is constant over the entire FinFET axis $x$ (solid red line), it can
be inferred that energy is conserved and that the GF+SSE model was
correctly implemented.
The following information can be
extracted from the data: (i) dissipated power (left plot), (ii) spectral
distribution of the electrical current (middle plot, red indicates
high current concentrations, green none), (iii) average crystal
temperature along the $x$-axis (middle plot), and (iv) heat
propagation map (right plot, red: heat flow towards left, blue: heat
flow towards right). The atomically resolved temperature of this
system was already presented in Fig.~\ref{fig:res}(d). It turns out
that most of the heat is generated close to the end of the transistor
channel ($x$=27nm). From there, it propagates towards the source
($x$=0) and drain ($x$=$L$) extensions, where it is absorbed by the
metal contacts. It can also be seen that the location with the highest
heat generation rate coincides with the maximum of the crystal
temperature ($T_{max}$). Both are situated in a region with a high
electric field, where electrons are prone to emit phonons.

Heat dissipation does not only affect the behavior of nano-scale
transistors, but also of a wide range of other nano-devices. For example,
Lithium-ion batteries (LIBs), under unfavorable operating conditions,
can overheat and endanger the security of the objects and persons
surrounding them. Understanding how heat is created in these energy
storage units (chemical reaction, Joule heating, other) and providing
design guidelines to reduce the temperature of their active region are
two objectives of utmost importance. Other nano-structures such as
non-volatile phase change random access memories (PCRAM) leverage
Joule heating to undergo a transition from an amorphous to a
crystalline phase. By better controlling this electro-thermal
phenomenon, it will be possible to obtain a more gradual change from
their high- to their low-resistance state such that PCRAMs can act as
solid-state synapses in ``non von Neumann'' neuromorphic computing
circuits. Current research in nano-transistors, LIBs, and PCRAM
cells, to cite a few applications, is expected to
benefit from the improvements that have been made to the
electro-thermal model of OMEN. Structure dimensions that were believed
to belong to the world of the imaginary are now accessible within
short turn-around times. Consequently, the proposed code
will allow to establish new bridges with experimental
groups working on nano-devices subject to strong heating
effects. 

\subsection{Data-Centric Parallel Programming}

The paper demonstrates how modifications to data movement alone can transform a 
complex, nonlinear solver to become communication efficient. Through modeling 
made possible by a data-centric intermediate representation, and graph transformations of 
the underlying macro- and micro- dataflow, this work is the first to introduce 
communication-avoiding principles to a full application. 

Because of the underlying retargetable SDFG representation, the solver runs on two different top-6 supercomputers efficiently, relying only on MPI and one external HPC library (BLAS) per-platform.
The SDFG was generated from a Python source code five times shorter than OMEN, and 
itself contains 2,015 nodes after transformations, created without modifying the original operations.
The resulting performance is \textit{two orders of magnitude} faster per-atom than the fine-tuned state of the art, which was recognized twice as a Gordon Bell finalist. 
This implies that overcoming scaling bottlenecks today requires reformulation and nontrivial decompositions, whose examination is facilitated by the data-centric paradigm.

\begin{acks}
This work was supported by the European Research Council
(ERC) under the European Union's Horizon 2020 programme
(grant agreement DAPP, No. 678880), by the MARVEL NCCR of the
Swiss National Science Foundation (SNSF), by the SNSF grant 175479
(ABIME), and by a grant from the Swiss National Supercomputing Centre, Project No.~s876. This work used resources of the Oak Ridge
Leadership Computing Facility, which is a DOE Office of Science User
Facility supported under Contract DE-AC05-00OR22725. The authors would
like to thank Maria Grazia Giuffreda, Nick Cardo (CSCS), Don Maxwell, Christopher Zimmer, and especially Jack Wells (ORNL) for access and support of the computational resources.
\end{acks}

\vspace{-0.5em}
\bibliographystyle{ACM-Reference-Format}
\bibliography{references}
    
\end{document}